\documentclass[12pt,twoside]{article}
\usepackage{graphicx}
\usepackage{amssymb}
\usepackage{amsmath}
\usepackage[dvipsnames]{xcolor}
\usepackage{orcidlink}
\usepackage[square,sort&compress,comma,numbers]{natbib}
\colorlet{Mycolor1}{green!10!red}
\begin{document}
\def\tf{\textbf}
\allowdisplaybreaks
\title{Multipartite Generalization of Geometric Measure of Discord}

\author{Ali Saif M. Hassan\orcidlink{⟨orcid⟩}$^{1,}$\footnote{Email: alisaif@amu.edu.ye and alisaif73@gmail.com} and Pramod S. Joag$^{2,}$ \footnote{Email: pramod@physics.unipune.ac.in}\\
$^1$ Department of Physics, University of Amran, Amran, Yemen.\\
$^2$ Department of Physics, University of Pune, Pune, India-411007.}
\maketitle

\begin{abstract}
Radhakrishnan \emph{et al.} [Phys. Rev. Lett. 124, 110401 (2020)] proposed a generalization of quantum discord to multipartite systems, which is consistent with the conventional deﬁnition of discord in bipartite systems, and derived explicit formulae for any multipartite state. These results are significant in capturing quantum correlations for multi-qubit systems.  We propose a generalization of the geometric measure of bipartite quantum discord to the multipartite case in the same manner. We ﬁnd generic forms of the generalization of the geometric measure of quantum discord in a general N-partite quantum state. Further, we obtain computable exact formulas for the generalization of the geometric measure of quantum discord in an N-qubit quantum state, which includes the results obtained in [arXiv:2104.12344]. \\

\noindent\textbf{ {Keywords}:} Quantum Discord,  Geometric Measure, Multipartite State, N-qubit State.\\
\noindent PACS numbers: 03.65.Ud;75.10.Pq;05.30.-d
\end{abstract}

\section{Introduction}

In quantum information theory, understanding and quantifying the different types of correlations in a quantum state is one of the fundamental problems that has led to intense research efforts over the last two decades. \cite{hor,guh}.
Correlations in quantum states, with far-reaching  implications for quantum information processing,
are usually studied in the entanglement-versus-separability framework \cite{hor,wer}.
However, it was shown that even some separable states contain nonclassical correlation, which can be
used to achieve some quantum tasks that cannot be accomplished by classical means \cite{knil,brau,benn,Meyer,biham,datt05,datt07}.
In the first two years of this century, another classification of quantum correlations based on quantum
measurements was proposed, based on the quantum-versus-classical model
of correlations \cite{pian}-\cite{lin08} and found to play an important role in quantum information theory.
It is considered to be an intrinsic type of non-classical correlation. These nonclassical correlations of bipartite states are measured by quantum discord \cite{olli, hend}, the discrepancy between quantum versions of two classically equivalent expressions for mutual information, or the minimized the difference between the quantum mutual information with and without the projective measurement performed on one of the subsystems in the bipartite system, which has generated increasing interest \cite{dill}-\cite{adesso}. The original deﬁnition of discord in terms of entropy is \cite{olli,hend}

\begin{equation}\label{e1}
QD(\rho) = \min_{\Pi^A} \{S_{B|\Pi^A}(\rho)-S_{B|A}(\rho)\},
\end{equation}
where $S_{B|A}(\rho)=S_{AB}(\rho)-S_A(\rho)$. The conditional entropy with measurement is deﬁned by

\begin{equation}\label{e2}
  S_{B|\Pi^A}(\rho)=\sum_j p^A_j S_{AB}(\Pi_j^A\rho\Pi_j^A/p^A_j),
\end{equation}
where $\Pi^A_j$ is a one-dimensional von Neumann projection operator on subsystem $A$ and
$p^A_j = Tr(\Pi^A_j\rho \Pi^A_j)$ is its probability. The discord is zero if and only if there is a measurement such that $\rho=\sum_j \Pi_j^A\rho\Pi_j^A.$\\

For a bipartite system, the distance-based approach introduced by Dakic et al. \cite{dakic}, called the geometric measure of discord is defined as:

\begin{equation}\label{e3}
GD(\rho)=\min_{\chi \in \Omega_0}||\rho-\chi||^2,
\end{equation}
\noindent where $\Omega_0$ denotes the set of zero-discord states and $||\rho-\chi||^2 := tr(\rho-\chi)^2$
is the square norm in the Hilbert-Schmidt space of linear operators acting on the state space of the system.

Dakic \emph{et al} \cite{dakic} also obtained an easily computable exact expression for the geometric measure
of quantum discord. For a two-qubit state $\rho$ expressed in its Bloch representation, this expression reads,

\begin{equation}\label{e4}
GD(\rho) = \frac{1}{4} (||x||^2 + ||T ||^2 - \lambda_{max}).
\end{equation}

\noindent Here $\vec{x}:= (x_1,x_2,x_3)^t$ and $\vec{y}:= (y_1,y_2,y_3)^t $ are coherent (column) vectors for single qubit reduced density operators,
$T = (t_{\alpha\beta})$ is the correlation matrix, and $\lambda_{max}$ is the largest eigenvalue of the
matrix $\vec{x}\vec{x}^t + T T^t .$ The norms of vectors and matrices are the Euclidean norms, for example, $||x||^2:=\sum_{\alpha} x^2_{\alpha}.$ Here and throughout this article, the superscript $t$ denotes the transpose of vectors and matrices, and by the norm of any tensor, we mean its Euclidean norm, that is, the square of the norm of a tensor is the sum of squares of its elements. This measure can be calculated analytically for arbitrary bipartite states  \cite{luo10}- \cite{rana12} as it is for general two-qubit states in Eq. (\ref{e4}).
Additionally, it has been demonstrated to exhibit operational signiﬁcance in speciﬁc quantum protocols (see Ref. \cite{dak12}).  {Despite} these important characteristics, geometric discord is recognized to be susceptible to the selection of distance measurements (see Ref. \cite{bell12}). However, as noted \cite{tuf12}-\cite{hu13}, it may increase under local operations on the unmeasured subsystem; the geometric discord as suggested in Ref. \cite{dakic} cannot be taken into consideration as an accurate measure of the quantum correlations.
Specifically, Piani \cite{pia12}  {has} demonstrated that the geometric discord is altered by a factor determined by the lack of purity of ancilla when a factorized local auxiliary state is introduced on the unmeasured party. In contrast, the entropic quantum discord is not affected by this issue.\\

By taking a look at several Schatten p-norms, Paula et al. \cite{pau13} presented the geometric discord and showed clearly that the 1-norm is the only p-norm that can establish a consistent quantum correlation measure.
Additionally, they gave an analytical formula for the 1-norm geometric discord, which turns out to be equivalent to the negativity of quantumness, by limiting the optimization to the tetrahedron of two-qubit Bell-diagonal states.\\

The process of determining geometric discord involves trace distance minimization, which might be difficult for general states. It is demonstrated that, {except for a} few families of states, it is nearly impossible to carry out operationally or analytically for qutrits and higher-dimensional bipartite systems \cite{lug17}. The full solution to the problem of determining the geometric discord trace norm for a general two-qubit state is described by Ługiewicz \cite{lug19}.\\

The Hilbert-Schmidt distance has the attractive characteristic of being reasonably simple to compute, despite its well-known abnormal  {behaviour} related to geometric discord. Even so, it is also useful since it can serve as evidence that quantum correlation exists in both bipartite and multipartite quantum systems, whereas quantum discord (QD) and the trace norm geometric measure lack calculable analytic formulas.\\

There have been many proposed attempts to generalize the original definition of discord to multipartite systems based on entropy, but these suggestions did not fit with the original definition of the bipartite system \cite{rul11}-\cite{cha11}.

Recently, Radhakrishnan \emph{et al.} \cite{rad} proposed the generalized quantum discord for multipartite quantum systems.
For multipartite systems with $N$ subsystems, $N-1$ local measurements are necessary to break all quantum correlations \cite{rul11, okr11}, with each successive measurement conditionally related to the previous.
The $N-1$-partite measurement is written as

\begin{equation}\label{e5}
\Pi^{A_1\cdots A_{N-1}}_{j_1 \cdots j_{N-1}}=\Pi^{A_1}_{j_1}\otimes \Pi^{A_2}_{j_2|j_1}\otimes \cdots \otimes \Pi^{A_{N-1}}_{j_{N-1}|j_1 \cdots j_{N-2}},
\end{equation}
\noindent where the $N$ subsystems are  {labelled} $A_i.$  Here the measurements take place in the order
 $A_1 \rightarrow A_2 \rightarrow \cdots \rightarrow A_{N-1}.$\\

The quantum discord of an $N$-partite state $\rho$ is deﬁned by
 \begin{equation}\label{e6}
   QD_{A_1;A_2;\cdots ;A_N}(\rho)=\min_{\Pi^{A_1\cdots A_{N-1}}}[-S_{A_2\cdots A_N|A_1}(\rho)+S_{A_2|\Pi^{A_1}}(\rho)+\cdots +S_{A_N|\Pi^{A_1\cdots A_{N-1}}}(\rho)],
 \end{equation}

\noindent where $S_{A_k|\Pi^{A_1\cdots A_{k-1}}}(\rho)=\sum_{j_1\cdots j_{k-1}} p^{(k-1)}_j S_{A_1\cdots A_k}(\Pi^{(k-1)}_j\rho \Pi^{(k-1)}_j/p^{(k-1)}_j)$
with $\Pi^{(k)}_j= \Pi^{A_1\cdots A_k}_{j_1 \cdots j_k}, \; p^{(k)}_j=tr(\Pi^{(k)}_j\rho \Pi^{(k)}_j).$\\

Guo \emph{et al.} and Zhu \emph{et al.} \cite{guo, zhu22} explored all the trade-off relations of multipartite quantum discord proposed very recently in \cite{rad} and  {showed} that the multipartite quantum discord is completely
monogamous provided that it does not increase under  {the} discard of subsystems. In addition, they explored all the trade-off relations for the global quantum discord proposed in \cite{rul11} and  {showed} that the global quantum discord is not completely monogamous. Li \emph{et al.} \cite{lib} evaluated analytically the quantum discord for a large family of multi-qubit states. Also, they investigated the dynamic  {behaviour} of quantum discord under decoherence. They proved that the phenomenon of frozen quantum discord does not exist under the phase ﬂip channel for the odd partite systems, while it can be found in the even partite systems.

For a geometric measure of quantum discord, Hassan and Joag \cite{hj12} have proposed its multipartite generalization.
The researchers found generic forms of quantum discord and total quantum correlations in a general $N$-partite quantum state under local unconditional successive measurements $\{\Pi^{A_1}_{j_1}\otimes \cdots \otimes\Pi^{A_N}_{j_N}\}$. Further, They obtained computable exact formulas for the geometric measure of quantum discord and total quantum correlations in an N-qubit quantum state.

The exact formulas for the $N$-qubit quantum state can be used to get experimental estimates of the quantum discord
and the total quantum correlation. A complete review of a distance-based approach can be found in ref. \cite{hu18}.

Recently, Zhu \emph{et al.} \cite{zhu} also, proposed the geometric measure of multipartite quantum discord and obtained results for a \textbf{family} of multi-qubit states. They investigated the dynamic  {behaviour} of geometric discord for the \textbf{family} of two-, three- and four-qubit states under phase noise acting on the ﬁrst qubit.\\

In this paper,
\begin{itemize}
\item We obtain a generic form for the generalization of  {the} geometric measure of discord $D_{A_1\cdots A_N}$ (corresponding to each successive measurement being conditionally related to the previous measurement) in {an} $N$-partite quantum state (Section 2, Theorem 1).

\item We give a formula for the generalization of {the} geometric measure of discord $D_{A_1 A_2 A_3}$ in a $3$-qubit quantum state, which is exactly computable (Section 3, Theorem 2).

\item We give a formula for the generalization of {the} geometric measure of discord $D_{A_1 A_2 A_3 A_4}$ in a $4$-qubit quantum state, which is exactly computable (Section 3, Theorem 3).

\item We give a formula for the generalization of {the} geometric measure of discord $D_{A_1 \cdots A_N}$ in {an} $N$-qubit quantum state, which is exactly computable (Section 3, Theorem 4).

\item We apply the exact formula for $D_{A_1 A_2 A_3}$ (obtained in Section 3) to some multiqubit states (Section 4).
\end{itemize}
Finally, we summarize in {Section} 5.

\section{Generalization of Geometric measure of Discord  in a $N$-Partite State}

Recall the definition of density matrix based on Hilbert-Schmitt space of liner operators \cite{hj12}.
Consider a multipartite system $\mathcal{H}=\mathcal{H}^{1}\otimes \mathcal{H}^{2} \otimes \cdots \otimes \mathcal{H}^{N}$
 with $dim(\mathcal{H}^{m}) = d_m, \; m=1,2,\cdots,N$. Let $L(\mathcal{H}^{m})$ be the Hilbert-Schmidt space of linear
operators on $\mathcal{H}^{m}$ with the Hilbert-Schmidt inner product
$$\langle X^{(m)}|Y^{(m)}\rangle := tr X^{(m)\dag} Y^{(m)}.$$
We can define  The Hilbert-Schmidt space $L(\mathcal{H}^{1}\otimes \mathcal{H}^{2} \otimes \cdots \otimes \mathcal{H}^{N})$
similarly.

Let $\{X^{(m)}_i : i = 1,2, . . . ,d_m^2,\; m=1,2,\cdots,N\}$ be set of Hermitian operators which constitute orthonormal
bases for $L(H^{m})$, then
$$tr X^{(m)}_i X^{(m)}_j = \delta_{ij},$$
and $\{X^{(1)}_{i_1}\otimes X^{(2)}_{i_2} \otimes \cdots \otimes X^{(N)}_{i_N}\}$ constitutes an orthonormal basis
for $L(\mathcal{H}^{1}\otimes \mathcal{H}^{2} \otimes \cdots \otimes \mathcal{H}^{N})$. In particular, any $n$-partite
state $\rho_{12\cdots N} \in L(\mathcal{H}^{1}\otimes \mathcal{H}^{2} \otimes \cdots \otimes \mathcal{H}^{N})$ can be
expanded as
\begin{equation}\label{e7}
\rho_{12\cdots N}= \sum_{i_1 i_2 \cdots i_N} c_{i_1 i_2 \cdots i_N} X^{(1)}_{i_1}\otimes X^{(2)}_{i_2} \otimes \cdots \otimes X^{(N)}_{i_N}\;;i_m  =1,\ldots,d_m^2\;;m=1,\ldots,N,
\end{equation}
with $\mathcal{C}=[c_{i_1 i_2 \cdots i_N}] = [tr(\rho_{12\cdots N} X^{(1)}_{i_1}\otimes X^{(2)}_{i_2} \otimes \cdots \otimes X^{(N)}_{i_N})]$
is an $N$-way array (tensor of order $N$) with size $d_1^2d_2^2\cdots d_N^2 .$

We can define the geometric measure of quantum discord for an $N$-partite quantum state corresponding to the von Neumann
measurement, each successive measurement is conditionally related to the previous measurement Eq.(\ref{e5}) as
\begin{equation}\label{e8}
GD_{A_1;A_2;\cdots ;A_N}(\rho)=\min_{\Pi^{A_1\cdots A_{N-1}}}||\rho_{12\cdots N}-\rho_{\Pi^{A_1\cdots A_{N-1}}}||^2 ,
\end{equation}
where the minimum is over the set of zero discord states \cite{rad},

$$\rho_{\Pi^{A_1\cdots A_{N-1}}}= \Pi^{A_1\cdots A_{N-1}}_{j_1 \cdots j_{N-1}} \rho \Pi^{A_1\cdots A_{N-1}}_{j_1 \cdots j_{N-1}}$$
$$=\sum_{j_1\cdots j_{N-1}} p^{A_1}_{j_1} p^{A_2}_{j_2|j_1}\cdots  p^{A_{N-1}}_{j_{N-1}|j_1\cdots j_{N-2}} |j_1\rangle \langle j_1|\otimes|j_2;j_1\rangle \langle j_2;j_1|\otimes \cdots$$
 \begin{equation}\label{e9}
  \otimes |j_{N=1};j_1\cdots j_{N-2}\rangle \langle j_{N-1};j_1\cdots j_{N-2}|\otimes \rho^{A_N}_{j_1\cdots j_{N-1}}
\end{equation}
where $j_k=1,2,\cdots,d_k.$ For the n-mode vector (matrix) product of a tensor, we refer the reader to the refs. \cite{kol06, lat00}.\\

We need to define a product of a tensor with a matrix, the n-mode product \cite{kol06, lat00}. The
$n$-mode (matrix) product of a tensor $\mathcal{Y}$ (of order $N$ and with dimension $J_1 \times J_2 \times \cdots \times J_N$ )
with a matrix $A$ with dimension $I \times J_n$ is denoted by $\mathcal{Y} \times_n A.$ The result is a tensor of size
$J_1 \times J_2 \times \cdots \times J_{n-1} \times I \times J_{n+1} \times \cdots \times J_N$ and is defined elementwise by

$$(\mathcal{Y} \times_n A)_{j_1 j_2\cdots j_{n-1} i j_{n+1}\cdots j_N} = \sum_{j_n=1}\mathcal{Y}_{j_1 j_2\cdots j_N} a_{ij_n}.$$

\textbf{Theorem 1}. Let $\rho_{12\cdots N}$ be a $N$-partite state defined by Eq.(\ref{e7}), then
\begin{equation}\label{e10}
GD_{A_1;A_2;\cdots ;A_N}(\rho_{12\cdots N})=||\mathcal{C}||^2-\max_{A_1,A_2^{(j_1)},\cdots,A_{N-1}^{(j_1j_2\cdots j_{N-2})}}||\mathcal{C}\times_1 A_1\times_2 A_2^{(j_1)}\times\cdots\times_{N-1}A_{N-1}^{(j_1j_2\cdots j_{N-2})}||^2,
\end{equation}
where $\mathcal{C}=[c_{i_1 i_2 \cdots i_N}]$ is defined via Eq.(\ref{e7}), the maximum is taken over all
$d_{k}\times d_{k}^2$-dimensional isometric matrices $A_k^{(j_1\cdots j_{k-1})}=[a_{li_k}^{(j_1\cdots j_{k-1})}],$ $A_k^{(j_1\cdots j_{k-1})}(A_k^{(j_1\cdots j_{k-1})})^t = I_{k},\; k=1,2,\cdots,N$
such that $a_{j_ki_k}^{(j_1\cdots j_{k-1})}=tr(|j_k;j_1\cdots j_{k-1}\rangle\langle j_k;j_1\cdots j_{k-1}| X^{(k)}_{i_k}), \;j_k=1,2,\ldots,d_{k};\;i_k=1,2,\ldots,d_{k}^2$
and $\{|j_k;j_1\cdots j_{k-1}\rangle\}$ is any orthonormal basis for $\mathcal{H}^{k}.$\\

\textbf{Proof:}
We expand the operator $|j_1\rangle\langle j_1|$ and $|j_k;j_1\cdots j_{k-1}\rangle \langle j_k;j_1\cdots j_{k-1}|,\;k=2,3,\cdots,N-1$
occurring in the expression for the zero discord state $\rho_{\Pi^{A_1\cdots A_{N-1}}}$ Eq.(\ref{e9})
in the orthonormal basis  $\{X_{i_{k}}^{(k)}\}$ in $L(\mathcal{H}^{k}),\; k=1,2,\cdots,N-1$ as
$$|j_1\rangle\langle j_1|=\sum_{i_1}^{d_1^2} a_{j_1i_1} X_{i_1}^{(1)},\; j_1=1,2,\cdots, d_1$$
\begin{equation}\label{e11}
|j_k;j_1\cdots j_{k-1}\rangle \langle j_k;j_1\cdots j_{k-1}|=\sum_{i_k=1}^{d_k^2} a_{j_ki_k}^{(j_1\cdots j_{k-1})} X_{i_{k}}^{(k)}, \; j_k=1,2,\cdots,d_{k},\; k=2,\cdots,N-1,
\end{equation}
with $$ a_{j_1i_1}=tr(|j_1\rangle \langle j_1| X_{i_1}^{(1)})=\langle j_1|X_{i_{1}}^{(1)}|j_1\rangle,$$
\begin{equation}\label{e12}
   a_{j_ki_k}^{(j_1\cdots j_{k-1})}=tr(|j_k;j_1\cdots j_{k-1}\rangle \langle j_k;j_1\cdots j_{k-1}| X_{i_k}^{(k)})=\langle j_k;j_1\cdots j_{k-1}|X_{i_{k}}^{(k)}|j_k;j_1\cdots j_{k-1}\rangle,
\end{equation}
$\{|j_k;j_1\cdots j_{k-1}\rangle \}$ being any conditional orthonormal basis in $\mathcal{H}^k.$ Clearly,
$\sum_{j_k=1}^{d_{k}}a_{j_ki_k}^{(j_1\cdots j_{k-1})} = tr X_{i_{k}}^{(k)}.$ Arranging the coefficients in a row vector as
$$ \vec{a}_{j_k}^{(j_1\cdots j_{k-1})}=(a_{j_k1}^{(j_1\cdots j_{k-1})},a_{j_k2}^{(j_1\cdots j_{k-1})},\cdots,a_{j_k d_{k}^2}^{(j_1\cdots j_{k-1})}),$$
we get, by the Parseval theorem of abstract Fourier transform,
\begin{equation}\label{e13}
  ||\vec{a}_{j_k}^{(j_1\cdots j_{k-1})}||^2=|||j_k;j_1\cdots j_{k-1}\rangle \langle j_k;j_1\cdots j_{k-1}|||^2=1.
\end{equation}
Where $||\vec{a}_{j_k}^{(j_1\cdots j_{k-1})}||^2= \sum_{i_{k}}(a_{j_ki_{k}}^{(j_1\cdots j_{k-1})})^2$. Moreover,
the orthonormality of $\{|j_k;j_1\cdots j_{k-1}\rangle\}$ implies that $\{\vec{a}_{j_k}^{(j_1\cdots j_{k-1})}\}$
is an orthonormal set of vectors, and therefore $A_k^{(j_1\cdots j_{k-1})} = [a_{j_ki_{k}}^{(j_1\cdots j_{k-1})}]$
is an isometry in the sense that $A_k^{(j_1\cdots j_{k-1})}(A_k^{(j_1\cdots j_{k-1})})^t = I_{k}.$

Similarly, we can expand the operator $p^{A_1}_{j_1} p^{A_2}_{j_2|j_1}\cdots  p^{A_{N-1}}_{j_{N-1}|j_1\cdots j_{N-2}}\rho^{A_N}_{j_1\cdots j_{N-1}}$
occurring in the expression for the zero discord state $\rho_{\Pi^{A_1\cdots A_{N-1}}}$ as
$$p^{A_1}_{j_1} p^{A_2}_{j_2|j_1}\cdots  p^{A_{N-1}}_{j_{N-1}|j_1\cdots j_{N-2}}\rho^{A_N}_{j_1\cdots j_{N-1}}
=\sum_{i_N}^{d_N^2}b_{i_N}^{(j_1\cdots j_{N-1})} X^{(N)}_{i_N},\;j_k=1,2,\cdots,d_{k},\; k=1,\cdots,N-1,$$
with $b_{i_N}^{(j_1\cdots j_{N-1})}= tr(p^{A_1}_{j_1} p^{A_2}_{j_2|j_1}\cdots  p^{A_{N-1}}_{j_{N-1}|j_1\cdots j_{N-2}}\rho^{A_N}_{j_1\cdots j_{N-1}} X_{i_N}^{(N)}).$\\

Then we have, using orthonormality of the basis,
\begin{equation}\label{e14}
  tr\rho_{\Pi^{A_1\cdots A_{N-1}}}^2=\sum_{j_1\cdots j_{n-1}}(p^{A_1}_{j_1})^2\cdots  (p^{A_{N-1}}_{j_{N-1}|j_1\cdots j_{N-2}})^2 tr(\rho^{A_N}_{j_1\cdots j_{N-1}})^2,
\end{equation}
and
\begin{equation}\label{e15}
\sum_{i_N}^{d_N^2} (b_{i_N}^{(j_1\cdots j_{n-1})})^2=(p^{A_1}_{j_1})^2\cdots  (p^{A_{N-1}}_{j_{N-1}|j_1\cdots j_{N-2}})^2 tr(\rho^{A_N}_{j_1\cdots j_{N-1}})^2.
\end{equation}
In view of Eqs.(\ref{e8}) and (\ref{e9}), the square norm distance between
$\rho_{12\cdots N}$ and $\rho_{\Pi^{A_1\cdots A_{N-1}}}$ can be evaluated (using the orthonormality of the bases
involved and Eq.(\ref{e11},\ref{e14}) as
\begin{eqnarray}
  ||\rho_{12\cdots N}-\rho_{\Pi^{A_1\cdots A_{N-1}}}||^2 &=& tr(\rho^2_{12\cdots N}-2 tr(\rho_{12\cdots N}\rho_{\Pi^{A_1\cdots A_{N-1}}})+tr\rho_{\Pi^{A_1\cdots A_{N-1}}}^2  \nonumber  \\
   &=& \sum_{i_1i_2\cdots  i_N} c^2_{i_1i_2\cdots  i_N} -2 \sum_{i_1i_2\cdots  i_N} c_{i_1i_2\cdots  i_N}\sum_{j_1\cdots j_{N-1}} p^{A_1}_{j_1} p^{A_2}_{j_2|j_1}\cdots  p^{A_{N-1}}_{j_{N-1}|j_1\cdots j_{N-2}}  \nonumber \\
  & & \times \langle j_1|X^{(1)}_{i_{1}}|j_1\rangle \cdots \langle j_{N-1};j_1\cdots j_{N-2}|X^{(N-1)}_{i_{N-1}}|j_{N-1};j_1\cdots j_{N-2}\rangle tr(\rho^{A_N}_{j_1\cdots j_{N-1}} X^{(N)}_{i_N})  \nonumber  \\
  & & +\sum_{j_1\cdots j_{N-1}} (p^{A_1}_{j_1})^2\cdots  (p^{A_{N-1}}_{j_{N-1}|j_1\cdots j_{N-2}})^2 tr(\rho^{A_N}_{j_1\cdots j_{N-1}})^2 \nonumber \\
   &=& ||\mathcal{C}||^2 - 2  \sum_{i_1i_2\cdots  i_N} c_{i_1i_2\cdots  i_N} \sum_{j_1\cdots j_{N-1}} a_{j_1i_1} a_{j_2i_2}^{(j_1)} \cdots a_{j_{N-1}i_{N-1}}^{(j_1\cdots j_{N-2})} b_{i_N}^{(j_1\cdots j_{N-1})} \nonumber  \\
   & &+\sum_{j_1\cdots j_{N-1}} \sum_{i_N}^{d_N^2} (b_{i_N}^{(j_1\cdots j_{N-1})})^2  \nonumber \\
   &=& ||\mathcal{C}||^2 -\sum_{j_1\cdots j_{N-1}} \sum_{i_N}^{d_N^2}\big{(} \sum_{i_1\cdots i_{N-1}} c_{i_1i_2\cdots  i_N}a_{j_1i_1} a_{j_2i_2}^{(j_1)} \cdots a_{j_{N-1}i_{N-1}}^{(j_1\cdots j_{N-2})} \big{)}^2  \nonumber \\
   & & +\sum_{j_1\cdots j_{N-1}}\sum_{i_N}^{d_N^2}\big{(}b_{i_N}^{(j_1\cdots j_{N-1})}-\sum_{i_1\cdots i_{N-1}} c_{i_1i_2\cdots  i_N}a_{j_1i_1} a_{j_2i_2}^{(j_1)} \cdots a_{j_{N-1}i_{N-1}}^{(j_1\cdots j_{N-2})}\big{)}^2. \nonumber
\end{eqnarray}
By choosing $b_{i_N}^{(j_1\cdots j_{n-1})}=\sum_{i_1\cdots i_{N-1}} c_{i_1i_2\cdots  i_N}a_{j_1i_1} a_{j_2i_2}^{(j_1)} \cdots a_{j_{N-1}i_{N-1}}^{(j_1\cdots j_{N-2})},$
(which can always be done because $\rho^{A_N}_{j_1\cdots j_{N-1}}$ is a set of arbitrary states and
$\{p^{A_1}_{j_1} p^{A_2}_{j_2|j_1}\cdots  p^{A_{N-1}}_{j_{N-1}|j_1\cdots j_{N-2}}\}$ is an arbitrary probability
distribution), the above equation reduces to $$||\rho_{12\cdots N}-\rho_{\Pi^{A_1\cdots A_{N-1}}}||^2=||\mathcal{C}||^2-||\mathcal{C}\times_1 A_1\times_2 A_2^{(j_1)}\times\cdots\times_{N-1}A_{N-1}^{(j_1j_2\cdots j_{N-2})}||^2.$$
Since the tensor $\mathcal{C}$ is determined by the state $\rho_{12\cdots N}$ via Eq.(\ref{e7}), we have,
using Eq.(\ref{e8}),
\begin{equation*}
GD_{A_1;A_2;\cdots ;A_N}(\rho_{12\cdots N})=||\mathcal{C}||^2-\max_{A_1,A_2^{(j_1)},\cdots,A_{N-1}^{(j_1j_2\cdots j_{N-2})}}||\mathcal{C}\times_1 A_1\times_2 A_2^{(j_1)}\times\cdots\times_{N-1}A_{N-1}^{(j_1j_2\cdots j_{N-2})}||^2,
\end{equation*}
where the maximum is taken over all $d_1\times d_1^2$-dimension $A_1$ and $d_{k}\times d_{k}^2$-dimensional
 $A_k^{(j_1\cdots j_{k-1})}=[a_{li_k}^{(j_1\cdots j_{k-1})}],\; k=2,\cdots,N-1$ isometric matrices specified in the
theorem, thus completing the proof.

\section{Generalization of Geometric measure for a $N$-qubit State}

Recall the definition of density matrix $\rho_{1\cdots N}$ in the Bloch representation \cite{hj12}.
We specialize in the $N$-qubit systems with states in
$\mathbb{C}^2\otimes\mathbb{C}^2\cdots \otimes \mathbb{C}^2$ ($N$ factors). We need the structure of the Bloch
representation of density operators, which can be briefly described as follows. Bloch representation of a density operator acting on the Hilbert space of a $d$-level quantum system $\mathbb{C}^d$ is given by
\begin{equation}\label{e16}
 \rho = \frac{1}{d} (I_d + \sum_{\alpha} s_{\alpha} {\tilde{\lambda}_{\alpha}}),
\end{equation}
where the components of the coherent vector $\vec{s},$ defined via Eq.(\ref{e8}), are given by
$s_{\alpha}=\frac{d}{2}tr(\rho {\tilde{\lambda}}_{\alpha}).$
Eq.(\ref{e8}) is the expansion of $\rho$ in the Hilbert-Schmidt basis $\{I_d,\tilde{\lambda}_{\alpha}; \alpha=1,2,\dots,d^2-1\}$
where $\tilde{\lambda}_{\alpha}$ are the traceless hermitian generators of $SU(d)$ satisfying
$tr(\tilde{\lambda}_{\alpha} \tilde{\lambda}_{\beta})=2\delta_{\alpha\beta}$  \cite{mahl}.\\

To give the Bloch representation of a density operator acting on the Hilbert
space $\mathbb{C}^{2} \otimes \mathbb{C}^{2} \otimes \cdots \otimes \mathbb{C}^{2}$
of a $n$-qubit quantum system, we introduce the following notation. We use $k_i \; (i=1,2,\cdots)$ to denote
a qubit chosen from $n$ qubits  {so} that  $k_i \; (i=1,2,\cdots)$  {takes} values in the set
$\mathcal{N}=\{1,2,\cdots,n\}$. Thus each $k_i$ is a variable taking values in $\mathcal{N}.$ The variables
$\alpha_{k_i}=1,2,3$ for a given $k_i$ span the set of generators of $SU(2)$ group (except identity) for the $k_i$th
qubit, namely the set of Pauli operators $\{\sigma_{1},\sigma_{2},\sigma_{3}\}$ for the $k_i$th qubit.
For two qubits $k_1$ and $k_2$ we define

$$\sigma^{(k_1)}_{\alpha_{k_1}} =  (I_2\otimes I_2\otimes \dots \otimes \sigma_{\alpha_{k_1}}\otimes I_2\otimes \dots \otimes I_2)$$
$$\sigma^{(k_2)}_{\alpha_{k_2}}  = (I_2\otimes I_2\otimes \dots \otimes \sigma_{\alpha_{k_2}}\otimes I_2\otimes \dots \otimes  I_2)$$
\begin{equation}\label{e17}
\sigma^{(k_1)}_{\alpha_{k_1}} \sigma^{(k_2)}_{\alpha_{k_2}}  = (I_2\otimes I_2\otimes \dots \otimes \sigma_{\alpha_{k_1}}\otimes
I_2\otimes \dots \otimes \sigma_{\alpha_{k_2}}\otimes I_2\otimes I_2) ,
\end{equation}
where  $\sigma_{\alpha_{k_1}}$ and $\sigma_{\alpha_{k_2}}$ occur at the $k_1$th and $k_2$th places (corresponding
to $k_1$th and $k_2$th qubits respectively) in the tensor product and are the $\alpha_{k_1}$th and  $\alpha_{k_2}$th
generators of $SU(2) ,\alpha_{k_1}=1,2,3\; \mbox{and} \; \alpha_{k_2}=1,2,3$ respectively. Then we can write,
for a $n$-qubit state $\rho_{12\cdots N} ,$

\begin{eqnarray}\label{e18}
 \rho_{12\cdots N} &=& \frac{1}{2^N} \{\otimes_m^N I_{d_m}+ \sum_{k_1 \in \mathcal{N}}\sum_{\alpha_{k_1}}s_{\alpha_{k_1}}\sigma^{(k_1)}_{\alpha_{k_1}} +\sum_{\{k_1,k_2\}}\sum_{\alpha_{k_1}\alpha_{k_2}}t_{\alpha_{k_1}\alpha_{k_2}}\sigma^{(k_1)}_{\alpha_{k_1}} \sigma^{(k_2)}_{\alpha_{k_2}}+\cdots   \nonumber \\
   & & +\sum_{\{k_1,k_2,\cdots,k_M\}}\sum_{\alpha_{k_1}\alpha_{k_2}\cdots \alpha_{k_M}}t_{\alpha_{k_1}\alpha_{k_2}\cdots \alpha_{k_M}}\sigma^{(k_1)}_{\alpha_{k_1}} \sigma^{(k_2)}_{\alpha_{k_2}}\cdots \sigma^{(k_M)}_{\alpha_{k_M}}+ \cdots \nonumber \\
   & & +\sum_{\alpha_{1}\alpha_{2}\cdots \alpha_{N}}t_{\alpha_{1}\alpha_{2}\cdots \alpha_{N}}\sigma^{(1)}_{\alpha_{1}} \sigma^{(2)}_{\alpha_{2}}\cdots \sigma^{(N)}_{\alpha_{N}}\},
\end{eqnarray}
where $\textbf{s}^{(k_1)}$ is a Bloch (coherent) vector corresponding to $k_1$th qubit,
$\mathbf{s}^{(k_1)} =[s_{\alpha_{k_1}}]_{\alpha_{k_1}=1}^{3} ,$ which is a tensor of order one defined by

\begin{equation}\label{e19}
  s_{\alpha_{k_1}}= tr[\rho \sigma^{(k_1)}_{\alpha_{k_1}}]= tr[\rho_{k_1} \sigma_{\alpha_{k_1}}],
 \end{equation}
where $\rho_{k_1}$ is the reduced density matrix for the $k_1$th qubit. Here $\{k_1,k_2,\cdots,k_M\},\; 2 \le M \le N,$
is a subset of $\mathcal{N}$ and can be chosen in $\binom{N}{M}$  ways, contributing $\binom{N}{M}$ terms in the
sum $\sum_{\{k_1,k_2,\cdots,k_M\}}$ in Eq.(\ref{e18}), each containing a tensor of order $M$.
The total number of terms in the Bloch representation of $\rho$ is $2^N$. We denote the tensors occurring in the sum
$\sum_{\{k_1,k_2,\cdots,k_M\}},\; (2 \le M \le N)$ by $T^{\{k_1,k_2,\cdots,k_M\}}=[t_{\alpha_{k_1}\alpha_{k_2}\cdots \alpha_{k_M}}]$
which  are defined by
\begin{equation}\label{e20}
 t_{\alpha_{k_1}\alpha_{k_2}\dots\alpha_{k_M}}= tr[\rho \sigma^{(k_1)}_{\alpha_{k_1}} \sigma^{(k_2)}_{\alpha_{k_2}}\cdots \sigma^{(k_M)}_{\alpha_{k_M}}]
 =tr[\rho_{k_1k_2\dots k_M} (\sigma_{\alpha_{k_1}}\otimes\sigma_{\alpha_{k_2}}\otimes\dots \otimes\sigma_{\alpha_{k_M}})]
\end{equation}
where $\rho_{k_1k_2\dots k_M}$ is the reduced density matrix for the subsystem $\{k_1 k_2\dots k_M\}$. We denote by $\mathcal{T}$ the tensor in the last term in Eq.(\ref{e18}).

In this article, we find the maximum in Eq.(\ref{e10}) for a $N$-qubit state $\rho_{12\cdots N}$ to obtain an exact
analytic formula, as in the two-qubit case (Eq.(\ref{e4})) \cite{dakic}.
We start with  {a} $3$-qubit state.\\

\textbf{Theorem 2}. Let $\rho_{123}$ be a $3$-qubit state defined by Eq.(\ref{e18}), then
\begin{eqnarray}\label{e21}
  GD_{A_1;A_2;A_3}(\rho_{123})  &=& \frac{1}{2^3}\left[||\vec{s}^{(1)}||^2+||\vec{s}^{(2)}||^2+||\mathcal{T}^{(12)}||^2+||\mathcal{T}^{(13)}||^2+||\mathcal{T}^{(23)}||^2+||\mathcal{T}||^2 \right.\nonumber \\
 & & \left.-\eta_{max}-\eta_{max}^{(1)}-\eta_{max}^{(2)}\right].
\end{eqnarray}
Here $\eta_{max},\;\eta_{max}^{(1)},$ and $\eta_{max}^{(2)}$  are the largest eigenvalues, corresponding to
$\hat{e},\;\hat{e}^{(1)},$ and $\hat{e}^{(2)}$ eigenvectors, of the matrix $G,\; G^{(1)},$ and $G^{(2)}$
which is a $3 \times 3$ real symmetric matrices, defined as

\begin{eqnarray} \label{e22}
G  &=& \vec{s}^{(1)}(\vec{s}^{(1)})^t+ T^{(13)}(T^{(13)})^t,  \nonumber \\
  G^{(1)} &=& \frac{1}{2}[\vec{s}^{(2)}(\vec{s}^{(2)})^t+\vec{y}^{(2)}(\vec{y}^{(2)})^t+ T^{(23)}(T^{(23)})^t+X^{(23)}(X^{(23)})^t   \nonumber  \\
   & & +\vec{s}^{(2)}(\vec{y}^{(2)})^t+\vec{y}^{(2)}(\vec{s}^{(2)})^t+T^{(23)}(X^{(23)})^t+X^{(23)}(T^{(23)})^t], \nonumber  \\
  G^{(2)} &=& \frac{1}{2}[\vec{s}^{(2)}(\vec{s}^{(2)})^t+\vec{y}^{(2)}(\vec{y}^{(2)})^t+ T^{(23)}(T^{(23)})^t+X^{(23)}(X^{(23)})^t   \nonumber \\
   & & -\vec{s}^{(2)}(\vec{y}^{(2)})^t-\vec{y}^{(2)}(\vec{s}^{(2)})^t-T^{(23)}(X^{(23)})^t-X^{(23)}(T^{(23)})^t],
 \end{eqnarray}
where $ \vec{y}^{(2)}_{i_2}=\sum_{i_1=1}^{3} \hat{e}_{i_1} t_{i_1i_2}^{(12)}$ and $X_{i_2i_3}^{(23)}=\sum_{i_1}^{3} \hat{e}_{i_1} t_{i_1i_2 i_3}.$\\

\textbf{Proof:}

Our goal is to get a closed-form expression for the term
$\max_{A_1,A_2^{(j_1)}}||\mathcal{C}\times_1 A_1\times_2 A_2^{(j_1)}||^2$ in Eq.(\ref{e10})
applied to an arbitrary state $\rho_{123}$ of a $3$-qubit system. The tensor $\mathcal{C}=[c_{i_1 i_2 i_3}]$ determined
by the $3$-qubit state $\rho_{123}$ via Eq.(\ref{e7}) has $i_m = 1,2,3,4;\; \; m=1,2,3,$ having $2^3$ elements in it.
The $2\times 4$ isometric matrices $A_1=[a_{j_1i_1}]$ have to satisfy $a_{j_1i_1}=tr(|j_1\rangle\langle j_1| X^{(1)}_{i_1}).$
In other words, the row vectors of $A_1$ must satisfy Eq.(\ref{e16}) for some single qubit pure state.
However, it is well known that every unit vector $\hat{s}$ in ${\mathbb{R}}^3$ satisfies Eq.(\ref{e16}) for some
single qubit pure state (which is not true for a higher dimensional system \cite{byr,kim}). Therefore, we can obtain
the required maximum  {overall} isometric $2\times 4$ matrices in the form obtained below (See Eq.s(\ref{e30},\ref{e31})).

We choose the orthonormal bases $\{X^{(m)}_{i_m}\},\; i_m=1,2,3,4$ in Eq.(\ref{e7}) as the generators of
$SU(2)^{\otimes m},\;m=1,2,3$ \cite{mahl}.

\begin{equation}\label{e23}
  X^{(m)}_1=\frac{1}{\sqrt{2}}I_2,
\end{equation}
 and

 \begin{equation}\label{e24}
   X^{(m)}_{i_m}= \frac{1}{\sqrt{2}}{\sigma}_{i_m-1},\; i_m=2,3,4;\;m=1,2,3
 \end{equation}
where ${\sigma}_{1,2,3}$ stand for the Pauli operators acting on the $m=1,2$th qubit.

 Since $tr{\sigma}_{\alpha_{k}}=0;\;\alpha_{k}=1,2,3$, we have,
\begin{equation*}
   \sum_{l=1}^{2} a_{li_{m}}=trX^{(m)}_{i_{m}}=\frac{1}{\sqrt{2}}tr{\sigma}_{i_{m}-1}=0,\; i_{m}=2,3,4.
\end{equation*}
Therefore,
\begin{equation}\label{e25}
  a_{2i_{1}}= - a_{1i_{1}}, \;\;\;
  a_{2i_{2}}^{(j_1)}= - a_{1i_{2}}^{(j_1)}, \; j_1=1,2; \;i_{1,2}=2,3,4.
\end{equation}
We now proceed to construct the $2\times 4$ matrices $A_1,\; A_2^{(j_1)}$ defined via Eq.(\ref{e12}). We will use
Eq.(\ref{e25}). The row vectors of $A_1,\; A_2^{(j_1)}$ are
$$\vec{a}_{j_1}=(a_{j_11},a_{j_12},a_{j_13},a_{j_14}); j_1=1,2.$$
$$\vec{a}_{j_2}^{(j_1)}=(a_{j_21}^{(j_1)},a_{j_22}^{(j_1)},a_{j_23}^{(j_1)},a_{j_24}^{(j_1)}); j_2=1,2.$$
Next, we define
\begin{equation}\label{e26}
 \hat{e}_{j_1}=\sqrt{2}(a_{j_12},a_{j_13},a_{j_14}),\;j_1=1,2, \; \; \;
 \hat{e}_{j_2}^{(j_1)}=\sqrt{2}(a_{j_22}^{(j_1)},a_{j_23}^{(j_1)},a_{j_24}^{(j_1)}),\;j_2=1,2,
 \end{equation}
and using Eq.(\ref{e25}), we get
\begin{equation}\label{e27}
 \hat{e}_2=- \hat{e}_1, \;\; \;
  \hat{e}_{2}^{(j_1)}=- \hat{e}_1^{(j_1)}, \; j_1=1,2.
 \end{equation}
We can prove
\begin{equation}\label{e28}
  ||\hat{e}_{j_1}||^2=1 ,\;\;j_1=1,2, \; \; \;
  ||\hat{e}_{j_2}^{(j_1)}||^2=1 ,\;\;j_2=1,2,
\end{equation}
using the condition Eq.(\ref{e13}),
$||\vec{a}_{j_1}||^2=\sum_{i_1=1}^{4} a_{j_1i_{1}}^2=1,\;\; ||\vec{a}_{j_2}^{(j_1)}||^2=\sum_{i_2=1}^{4} (a_{j_2i_2}^{(j_1)})^2=1,$
and $a_{j_11}= tr(|j_1\rangle \langle j_1| X^{(1)}_1)=\frac{1}{\sqrt{2}}, \; \; a_{j_21}^{(j_1)}=tr(|j_2;j_1\rangle \langle j_2;j_1| X^{(2)}_1)=\frac{1}{\sqrt{2}}.$

We can now construct the row vectors of $2\times 4$ matrices $A_1,\;A_2^{(j_1)}$, using Eq.(\ref{e26}) and
Eq.(\ref{e27}), with defining $\hat{e}_1=\hat{e}$ and $\hat{e}_1^{(j_1)}=\hat{e}^{(j_1)},\; j_1=1,2.$
\begin{equation}\label{e29}
  \vec{a}_1=\frac{1}{\sqrt{2}} (1, \hat{e}),  \; \; \;
  \vec{a}_2=\frac{1}{\sqrt{2}} (1, -\hat{e}),
  \end{equation}
\begin{equation}\label{e30}
  \vec{a}_1^{(j_1)}=\frac{1}{\sqrt{2}} (1, \hat{e}^{(j_1)}),  \; \; \;
  \vec{a}_2^{(j_1)}=\frac{1}{\sqrt{2}} (1, -\hat{e}^{(j_1)}),
\end{equation}
The matrices $A_1,$  {and} $A_2^{(j_1)}$ for 1st and 2nd systems respectively are, in terms of the row vectors defined above,

\begin{displaymath}
A_1=\frac{1}{\sqrt{2}}
\left(\begin{array}{cc}
1 & \hat{e}\\
1 & -\hat{e}\\
\end{array}\right),
\end{displaymath}
and

\begin{displaymath}
A_2^{(j_1)}=\frac{1}{\sqrt{2}}
\left(\begin{array}{cc}
1 & \hat{e}^{(j_1)}\\
1 & -\hat{e}^{(j_1)}\\
\end{array}\right).
\end{displaymath}
The norm of the tensor $\mathcal{C}$ can be expressed in terms of the norms of the tensors defining $\rho_{123}$ by
using the equivalence of the definitions of $\rho_{123}$ given in Eq.(\ref{e7}) and Eq.(\ref{e18}) as

\begin{equation}\label{e31}
  ||\mathcal{C}||^2  =  \frac{1}{2^3}\left[1+||\vec{s}^{(1)}||^2+ ||\vec{s}^{(2)}||^2+ ||\vec{s}^{(3)}||^2+||T^{(12)}||^2+||T^{(13)}||^2+||T^{(23)}||^2+||\mathcal{T}||^2\right].
\end{equation}
To get the norm of $\mathcal{C}\times_1 A_1\times_2 A_2^{(j_1)}$ we use its elementwise definition,
\begin{equation}\label{e32}
  (\mathcal{C}\times_1 A_1\times_2 A_2^{(j_1)})_{j_1j_2i_3}  =
   \sum_{i_1i_2} c_{i_1i_2 i_3} a_{j_1i_1}a_{j_2 i_2}^{(j_1)},  \; j_1=1,2, \; j_2=1,2,
\end{equation}
the equivalence of the definitions of $\rho_{123}$ given in Eq.(\ref{e7}) and Eq.(\ref{e18}) and the elements of
$A_1,\; A_2^{(j_1)}$ given by Eq.s(\ref{e29},\ref{e30}). The result is

\begin{eqnarray}\label{e33}
  ||\mathcal{C}\times_1 A_1 \times_2 A_2^{(j_1)}||^2 &=& \frac{1}{2^3}\left\{1+||\vec{s}^{(3)}||^2 + \hat{e} \; \vec{s}^{(1)}(\vec{s}^{(1)})^t \hat{e}^t +\hat{e} T^{(13)}(T^{(13)})^t \hat{e}^t+\frac{1}{2}\left[\hat{e}^{(1)}\vec{s}^{(2)}(\vec{s}^{(2)})^t (\hat{e}^{(1)})^t \right. \right. \nonumber \\
   & & \left.+\hat{e}^{(1)} T^{(23)}(T^{(23)})^t (\hat{e}^{(1)})^t+ \hat{e}^{(1)}\vec{y}^{(2)}(\vec{y}^{(2)})^t (\hat{e}^{(1)})^t +\hat{e}^{(1)} X^{(23)}(X^{(23)})^t (\hat{e}^{(1)})^t \right.\nonumber \\
   & & + \left.\left.\hat{e}^{(1)}\vec{y}^{(2)}(\vec{s}^{(2)})^t (\hat{e}^{(1)})^t+\hat{e}^{(1)} X^{(23)}(T^{(23)})^t (\hat{e}^{(1)})^t+\hat{e}^{(1)}\vec{s}^{(2)}(\vec{y}^{(2)})^t (\hat{e}^{(1)})^t \right.\right. \nonumber \\
   & & \left. \left.  +\hat{e}^{(1)} T^{(23)}(X^{(23)})^t (\hat{e}^{(1)})^t \right]+\frac{1}{2}\left[\hat{e}^{(2)}\vec{s}^{(2)}(\vec{s}^{(2)})^t (\hat{e}^{(2)})^t +\hat{e}^{(2)} T^{(23)}(T^{(23)})^t (\hat{e}^{(2)})^t   \right.\right. \nonumber \\
   & & \left. \left.  + \hat{e}^{(2)}\vec{y}^{(2)}(\vec{y}^{(2)})^t (\hat{e}^{(2)})^t +\hat{e}^{(2)} X^{(23)}(X^{(23)})^t (\hat{e}^{(2)})^t - \hat{e}^{(2)}\vec{y}^{(2)}(\vec{s}^{(2)})^t (\hat{e}^{(2)})^t  \right. \right. \nonumber \\
   & &  \left. \left. -\hat{e}^{(2)} X^{(23)}(T^{(23)})^t (\hat{e}^{(2)})^t-\hat{e}^{(2)}\vec{s}^{(2)}(\vec{y}^{(2)})^t (\hat{e}^{(2)})^t -\hat{e}^{(2)} T^{(23)}(X^{(23)})^t (\hat{e}^{(2)})^t \right] \right\} \nonumber \\
\end{eqnarray}
or,
\begin{eqnarray}\label{e34}
  ||\mathcal{C}\times_1 A_1 \times_2 A_2^{(j_1)}||^2 &=& \frac{1}{2^3}\left\{1+||\vec{s}^{(3)}||^2 + \hat{e} \; [\vec{s}^{(1)}(\vec{s}^{(1)})^t  + T^{(13)}(T^{(13)})^t]\hat{e}^t+\frac{1}{2}\hat{e}^{(1)}\left[\vec{s}^{(2)}(\vec{s}^{(2)})^t  \right. \right. \nonumber \\
   & & \left. \left.+ T^{(23)}(T^{(23)})^t + \vec{y}^{(2)}(\vec{y}^{(2)})^t  + X^{(23)}(X^{(23)})^t  \right.  \right.\nonumber \\
   & & + \left.\left.\vec{y}^{(2)}(\vec{s}^{(2)})^t + X^{(23)}(T^{(23)})^t +\vec{s}^{(2)}(\vec{y}^{(2)})^t \right.\right. \nonumber \\
   & & \left. \left.  + T^{(23)}(X^{(23)})^t  \right](\hat{e}^{(1)})^t
   +\frac{1}{2}\hat{e}^{(2)}\left[\vec{s}^{(2)}(\vec{s}^{(2)})^t  + T^{(23)}(T^{(23)})^t  \right. \right. \nonumber \\
   & & \left. \left.  +\vec{y}^{(2)}(\vec{y}^{(2)})^t + X^{(23)}(X^{(23)})^t - \vec{y}^{(2)}(\vec{s}^{(2)})^t \right. \right. \nonumber \\
   & &  \left. \left. - X^{(23)}(T^{(23)})^t -\vec{s}^{(2)}(\vec{y}^{(2)})^t  - T^{(23)}(X^{(23)})^t \right] (\hat{e}^{(2)})^t \right\}\nonumber \\
\end{eqnarray}
where $$\vec{y}^{(2)}_{i_2}=\sum_{i_1=1}^{3}\hat{e}_{i_1} t_{i_1i_2}^{(12)},\; \; X_{i_2i_3}^{(23)}=\sum_{i_1=1}^{3}\hat{e}_{i_1} t_{i_1i_2i_3}^{(123)}$$
as in the statement of the theorem.

Let us identify the expressions in square brackets in Eq.(\ref{e34}), by the  $(3\times 3)$ real symmetric  matrices  as

\begin{eqnarray*}
 G &=& \vec{s}^{(1)}(\vec{s}^{(1)})^t +T^{(13)}(T^{(13)})^t \\
G^{(1)} &=& \frac{1}{2}[ \vec{s}^{(2)}(\vec{s}^{(2)})^t+ T^{(23)}(T^{(23)})^t + \vec{y}^{(2)}(\vec{y}^{(2)})^t+ X^{(23)}(X^{(23)})^t \\
   & & + \vec{y}^{(2)}(\vec{s}^{(2)})^t + X^{(23)}(T^{(23)})^t+\vec{s}^{(2)}(\vec{y}^{(2)})^t+ T^{(23)}(X^{(23)})^t, \\
 G^{(2)} &=& \frac{1}{2}[ \vec{s}^{(2)}(\vec{s}^{(2)})^t+ T^{(23)}(T^{(23)})^t + \vec{y}^{(2)}(\vec{y}^{(2)})^t+ X^{(23)}(X^{(23)})^t \\
    & &- \vec{y}^{(2)}(\vec{s}^{(2)})^t - X^{(23)}(T^{(23)})^t -\vec{s}^{(2)}(\vec{y}^{(2)})^t- T^{(23)}(X^{(23)})^t,
\end{eqnarray*}
Thus, we get
\begin{equation}\label{e35}
||\mathcal{C}\times_1 A_1 \times_2 A_2^{(j_1)}||^2= \frac{1}{2^3}\left\{1+||\vec{s}^{(3)}||^2 + \hat{e} G \hat{e}^t + \hat{e}^{(1)} G^{(1)} (\hat{e}^{(1)})^t + \hat{e}^{(2)} G^{(2)} (\hat{e}^{(2)})^t \right\}.
\end{equation}
In Eq.(\ref{e35}) the three last terms depend on matrices $A_1,$  {and} $A_2^{(j_1)}$ while others are determined by the state
$\rho_{123}.$ Therefore, to maximize $||\mathcal{C}\times_1 A_1 \times_2 A_2^{(j_1)}||^2$ we take $\hat{e},\; \hat{e}^{(1)},\; \hat{e}^{(2)}$
to be the eigenvectors of $G, \; G^{(1)},\; G^{(2)}$ corresponding to its largest eigenvalues
$\eta_{max},\;\eta_{max}^{(1)}, \; \eta_{max}^{(2)}$ respectively,  so that

\begin{equation}\label{e36}
\max_{A_1,A_2^{(j_1)}}||\mathcal{C}\times_1 A_1 \times_2 A_2^{(j_1)}||^2 = \frac{1}{2^3}\left\{1+||\vec{s}^{(3)}||^2 + \eta_{max}+\eta_{max}^{(1)}+\eta_{max}^{(2)} \right\}.
\end{equation}
Finally, Eq.(\ref{e36}), Eq.(\ref{e31}) and  Eq.(\ref{e10}) together imply

\begin{eqnarray}\label{e37}
GD_{A_1;A_2;A_3}(\rho_{123})&=& \frac{1}{2^3}\left\{||\vec{s}^{(1)}||^2+||\vec{s}^{(2)}||^2+||T^{(12)}||^2+||T^{(13)}||^2
  +||T^{(23)}||^2+||\mathcal{T}||^2 \right. \nonumber \\
 & &\left. -\eta_{max}-\eta_{max}^{(1)}-\eta_{max}^{(2)}\right\},
\end{eqnarray}
where $\eta_{max},\;\eta_{max}^{(1)}, \; \eta_{max}^{(2)}$ are the largest eigenvalues of matrices
$G, \; G^{(1)},\; G^{(2)}$ respectively, thus completing the proof.\\
For the $4$-qubit state, we state \\

\textbf{Theorem 3}. Let $\rho_{1234}$ be a $4$-qubit state defined by Eq.(\ref{e18}), then
\begin{eqnarray}\label{e38}
   GD_{A_1;A_2;A_3;A_4}(\rho_{1234}) &=& \frac{1}{2^4} \left[ ||\vec{s}^{(1)}||^2+||\vec{s}^{(2)}||^2+||\vec{s}^{(3)}||^2+\sum_{1\leq k <l\leq4}||T^{(kl)}||^2+\sum_{1\leq k< l <n \leq 4}||T^{(kln)}||^2 \right. \nonumber \\
   & &  \left. +||\mathcal{T}||^2-\eta_{max}  -\sum_{j_1=1}^{2}\eta_{max}^{(j_1)}
  -\sum_{j_1=1}^{2} \sum_{j_2=1}^{2}\eta_{max}^{(j_1j_2)}\right].
 \end{eqnarray}
Here $\eta_{max},\;\eta_{max}^{(j_1)},\;\eta_{max}^{(j_1j_2)}$  are the largest eigenvalues, corresponding to
$\hat{e},\;\hat{e}^{(j_1)},\; \hat{e}^{(j_1j_2)}$ eigenvectors, of the matrix
$G,\; G^{(j_1)},\;  G^{(j_1j_2)},\; \; j_1,j_2=1,2$ which is a $3 \times 3$ real symmetric matrices, defined as

\begin{eqnarray} \label{e39}
G  &=& \vec{s}^{(1)}(\vec{s}^{(1)})^t+ T^{(14)}(T^{(14)})^t,  \nonumber \\
  G^{(j_1)} &=& \frac{1}{2}[\vec{s}^{(2)}(\vec{s}^{(2)})^t+\vec{y}^{(2)}(\vec{y}^{(2)})^t+ T^{(24)}(T^{(24)})^t+X^{(24)}(X^{(24)})^t   \nonumber  \\
   & & +(-1)^{j_1-1}(\vec{s}^{(2)}(\vec{y}^{(2)})^t+\vec{y}^{(2)}(\vec{s}^{(2)})^t+T^{(24)}(X^{(24)})^t+X^{(24)}(T^{(24)})^t)], \; j_1=1,2 \nonumber  \\
  G^{(j_1j_2)} &=& \frac{1}{4}[\vec{s}^{(3)}(\vec{s}^{(3)})^t+ T^{(34)}(T^{(34)})^t+\vec{y}^{(3)}(\vec{y}^{(3)})^t
  +X^{(34)}(X^{(34)})^t+ \vec{y}^{(3,j_1)}(\vec{y}^{(3,j_1)})^t +X^{(34,j_1)}(X^{(34,j_1)})^t    \nonumber \\
   & & + \vec{z}^{(3,j_1)}(\vec{z}^{(3,j_1)})^t +W^{(34,j_1)}(W^{(34,j_1)})^t+(-1)^{j_1-1} [\vec{s}^{(3)} (\vec{y}^{(3)})^t + T^{(34)}(X^{(34)})^t \nonumber  \\
   & & +\vec{y}^{(3)}(\vec{s}^{(3)})^t+X^{(34)}(T^{(34)})^t +\vec{y}^{(3,j_1)}(\vec{z}^{(3,j_1)})^t +X^{(34,j_1)}(W^{(34,j_1)})^t+\vec{z}^{(3,j_1)}(\vec{y}^{(3,j_1)})^t\nonumber  \\
   & &+ W^{(34,j_1)}(X^{(34,j_1)})^t]+(-1)^{j_2-1} [\vec{s}^{(3)} (\vec{y}^{(3,j_1)})^t + T^{(34)}(X^{(34,j_1)})^t \nonumber  \\
   & & +\vec{y}^{(3,j_1)}(\vec{s}^{(3)})^t+X^{(34,j_1)}(T^{(34)})^t +\vec{y}^{(3)}(\vec{z}^{(3,j_1)})^t +X^{(34)}(W^{(34,j_1)})^t+\vec{z}^{(3,j_1)}(\vec{y}^{(3)})^t  \nonumber  \\
   & &+ W^{(34,j_1)}(X^{(34)})^t]+(-1)^{j_1-1}(-1)^{j_2-1}[\vec{s}^{(3)}(\vec{z}^{(3,j_1)})^t+T^{(34)}(W^{(34,j_1)})^t  +\vec{z}^{(3,j_1)}(\vec{s}^{(3)})^t   \nonumber  \\
   & &+W^{(34,j_1)}(T^{(34)})^t+\vec{y}^{(3)} (\vec{y}^{(3,j_1)})^t + X^{(34)}(X^{(34,j_1)})^t+\vec{y}^{(3,j_1)}(\vec{y}^{(3)})^t+X^{(34,j_1)}(X^{(34)})^t],
 \end{eqnarray}
where $ \vec{y}^{(2)}_{i_2}=\sum_{i_1=1}^{3} \hat{e}_{i_1} t_{i_1i_2}^{(12)},\; X_{i_2i_4}^{(24)}=\sum_{i_1}^{3} \hat{e}_{i_1} t_{i_1i_2 i_4}^{(124)},\;\; \vec{y}^{(3)}_{i_3}=\sum_{i_1=1}^{3} \hat{e}_{i_1} t_{i_1i_3}^{(13)},$\\
$\; \; X_{i_3i_4}^{(34)}=\sum_{i_1}^{3} \hat{e}_{i_1} t_{i_1i_3 i_4}^{(134)},$
and \;
$ \vec{y}^{(3,j_1)}_{i_3}=\sum_{i_2=1}^{3} \hat{e}_{i_2}^{(j_1)} t_{i_2i_3}^{(23)},\; X_{i_3i_4}^{(34,j_1)}=\sum_{i_2}^{3} \hat{e}_{i_2}^{(j_1)} t_{i_2i_3 i_4}^{(234)},$\\ $\vec{z}^{(3,j_1)}_{i_3}=\sum_{i_1=1}^{3} \sum_{i_2=1}^{3} \hat{e}_{i_1} \hat{e}_{i_2}^{(j_1)} t_{i_1i_2i_3}^{(123)},\; \; W_{i_3i_4}^{(34,j_1)}=\sum_{i_1}^{3} \sum_{i_2=1}^{3}\hat{e}_{i_1}\hat{e}_{i_2}^{(j_1)} t_{i_1i_2i_3 i_4}^{(1234)}.$\\

\textbf{Proof:}
Following the proof of theorem 2, equation (\ref{e25}) becomes
\begin{equation}\label{e40}
  a_{2i_{1}}= - a_{1i_{1}}, \;\;\;
  a_{2i_{2}}^{(j_1)}= - a_{1i_{2}}^{(j_1)},
  a_{2i_{3}}^{(j_1j_2)}= - a_{1i_{3}}^{(j_1j_2)},  \; j_k=1,2; \;i_{k}=2,3,4; \; k=1,2,3.
\end{equation}
We now proceed to construct the $2\times 4$ matrices  $A_1,\; A_2^{(j_1)},\;A_3^{(j_1j_2)}$ defined via Eq.(\ref{e12}).
We will use Eq.(\ref{e40}). The row vectors of $A_1,\; A_2^{(j_1)},\;A_3^{(j_1j_2)}$ are
$$\vec{a}_{j_1}=(a_{j_11},a_{j_12},a_{j_13},a_{j_14}); j_1=1,2.$$
$$\vec{a}_{j_2}^{(j_1)}=(a_{j_21}^{(j_1)},a_{j_22}^{(j_1)},a_{j_23}^{(j_1)},a_{j_24}^{(j_1)}); j_2=1,2.$$
$$\vec{a}_{j_3}^{(j_1j_2)}=(a_{j_31}^{(j_1j_2)},a_{j_32}^{(j_1j_2)},a_{j_33}^{(j_1j_2)},a_{j_34}^{(j_1j_2)}); j_3=1,2.$$
Next, we define
\begin{equation*}
 \hat{e}_{j_1}=\sqrt{2}(a_{j_12},a_{j_13},a_{j_14}),\;j_1=1,2, \; \; \;
 \hat{e}_{j_2}^{(j_1)}=\sqrt{2}(a_{j_22}^{(j_1)},a_{j_23}^{(j_1)},a_{j_24}^{(j_1)}),\;j_2=1,2,
 \end{equation*}
 \begin{equation}\label{e41}
 \hat{e}_{j_3}^{(j_1j_2)}=\sqrt{2}(a_{j_32}^{(j_1j_2)},a_{j_33}^{(j_1j_2)},a_{j_34}^{(j_1j_2)}),\;j_3=1,2,
  \end{equation}
and using Eq.(\ref{e40}), we get
\begin{equation}\label{e42}
 \hat{e}_2=- \hat{e}_1, \;\; \;
  \hat{e}_{2}^{(j_1)}=- \hat{e}_1^{(j_1)}, \; j_1=1,2. \; \;
 \hat{e}_{2}^{(j_1j_2)}=- \hat{e}_1^{(j_1j_2)}, \; j_1,j_2=1,2.
 \end{equation}
We can prove
\begin{equation}\label{e43}
  ||\hat{e}_{j_1}||^2=1 ,\;\;j_1=1,2, \; \; \;
  ||\hat{e}_{j_2}^{(j_1)}||^2=1 ,\;\;j_2=1,2,
 ||\hat{e}_{j_3}^{(j_1j_2)}||^2=1 ,\;\;j_3=1,2,
\end{equation}
using the condition Eq.(\ref{e13}),
$||\vec{a}_{j_1}||^2=\sum_{i_1=1}^{4} a_{j_1i_{1}}^2=1,\;\; ||\vec{a}_{j_2}^{(j_1)}||^2=\sum_{i_2=1}^{4} (a_{j_2i_2}^{(j_1)})^2=1,\;\; ||\vec{a}_{j_3}^{(j_1j_2)}||^2=\sum_{i_3=1}^{4} (a_{j_3i_3}^{(j_1j_2)})^2=1,$
and $a_{j_11}= tr(|j_1\rangle \langle j_1| X^{(1)}_1)=\frac{1}{\sqrt{2}}, \; \; a_{j_21}^{(j_1)}=tr(|j_2;j_1\rangle \langle j_2;j_1| X^{(2)}_1)=\frac{1}{\sqrt{2}},\; \; a_{j_31}^{(j_1j_2)}=tr(|j_3;j_1j_2\rangle \langle j_3;j_1j_2| X^{(3)}_1)=\frac{1}{\sqrt{2}}.$

We can now construct the row vectors of $2\times 4$ matrices $A_1,\;A_2^{(j_1)},\;A_3^{(j_1j_2)}$, using Eq.(\ref{e41})
and Eq.(\ref{e42}), with defining $\hat{e}_1=\hat{e},\;\hat{e}_1^{(j_1)}=\hat{e}^{(j_1)},\; j_1=1,2,$ and
$\hat{e}_1^{(j_1j_2)}=\hat{e}^{(j_1j_2)},\; j_1,j_2=1,2.$

\begin{equation}\label{e44}
  \vec{a}_1=\frac{1}{\sqrt{2}} (1, \hat{e}),  \; \; \;
  \vec{a}_2=\frac{1}{\sqrt{2}} (1, -\hat{e}),
  \end{equation}
\begin{equation}\label{e45}
  \vec{a}_1^{(j_1)}=\frac{1}{\sqrt{2}} (1, \hat{e}^{(j_1)}),  \; \; \;
  \vec{a}_2^{(j_1)}=\frac{1}{\sqrt{2}} (1, -\hat{e}^{(j_1)}),
\end{equation}

\begin{equation}\label{e46}
  \vec{a}_1^{(j_1j_2)}=\frac{1}{\sqrt{2}} (1, \hat{e}^{(j_1j_2)}),  \; \; \;
  \vec{a}_2^{(j_1j_2)}=\frac{1}{\sqrt{2}} (1, -\hat{e}^{(j_1j_2)}),
\end{equation}
The matrices $A_1,\; A_2^{(j_1)},$  {and} $ A_3^{(j_1j_2)}$ for 1st, 2nd and 3rd systems respectively are, in terms of  {their} row
vectors defined above,

\begin{equation*}
A_1=\frac{1}{\sqrt{2}}
\left(\begin{array}{cc}
1 & \hat{e}\\
1 & -\hat{e}\\
\end{array}\right),
\end{equation*}
and

\begin{equation*}
A_2^{(j_1)}=\frac{1}{\sqrt{2}}
\left(\begin{array}{cc}
1 & \hat{e}^{(j_1)}\\
1 & -\hat{e}^{(j_1)}\\
\end{array}\right).
\end{equation*}
\begin{equation*}
A_3^{(j_1j_2)}=\frac{1}{\sqrt{2}}
\left(\begin{array}{cc}
1 & \hat{e}^{(j_1j_2)}\\
1 & -\hat{e}^{(j_1j_2)}\\
\end{array}\right).
\end{equation*}
The norm of the tensor $\mathcal{C}$ can be expressed in terms of the norms of the tensors defining $\rho_{1234}$ by
using the equivalence of the definitions of $\rho_{1234}$ given in Eq.(\ref{e7}) and Eq.(\ref{e18}) as

\begin{eqnarray}\label{e47}
  ||\mathcal{C}||^2 &=&  \frac{1}{2^4}\left[1+||\vec{s}^{(1)}||^2+ ||\vec{s}^{(2)}||^2+ ||\vec{s}^{(3)}||^2+||\vec{s}^{(4)}||^2+||T^{(12)}||^2+||T^{(13)}||^2+||T^{(14)}||^2+||T^{(23)}||^2 \right. \nonumber \\
   & &\left. +||T^{(24)}||^2+||T^{(34)}||^2+||T^{(123)}||^2+||T^{(124)}||^2+||T^{(134)}||^2
  +||T^{(234)}||^2+||\mathcal{T}||^2\right].
\end{eqnarray}
In order to get the norm of $\mathcal{C}\times_1 A_1\times_2 A_2^{(j_1)}\times_3 A_2^{(j_1j_2)}$ we use its elementwise
definition,
\begin{equation}\label{e48}
  (\mathcal{C}\times_1 A_1\times_2 A_2^{(j_1)}\times_3 A_3^{(j_1j_2)})_{j_1j_2j_3i_4}  =
   \sum_{i_1i_2i_3} c_{i_1i_2 i_3i_4} a_{j_1i_1}a_{j_2 i_2}^{(j_1)} a_{j_3 i_3}^{(j_1j_2)},  \; j_1=1,2, \; j_2=1,2,\; j_3=1,2,
\end{equation}
the equivalence of the definitions of $\rho_{1234}$ given in Eq.(\ref{e7}) and Eq.(\ref{e18}) and the elements of
$A_1,\; A_2^{(j_1)},\; A_3^{(j_1j_2)}$ given by Eq.s(\ref{e44},\;\ref{e45},\, \ref{e46}). The result is

\begin{eqnarray}\label{e49}
  ||\mathcal{C}\times_1 A_1 \times_2 A_2^{(j_1)}\times_3 A_3^{(j_1j_2)}||^2 &=& \frac{1}{2^4}\left\{1+||\vec{s}^{(4)}||^2 + \hat{e} \; \vec{s}^{(1)}(\vec{s}^{(1)})^t \hat{e}^t +\hat{e} T^{(14)}(T^{(14)})^t \hat{e}^t  \right.  \nonumber \\
  & &\left. +\sum_{j_1=1}^{2}\frac{1}{2}\left[\hat{e}^{(j_1)}(\vec{s}^{(2)}(\vec{s}^{(2)})^t+T^{(24)}(T^{(24)})^t) (\hat{e}^{(j_1)})^t \right. \right. \nonumber \\
  & &+\left. \left. \hat{e}^{(j_1)}(\vec{y}^{(2)}(\vec{y}^{(2)})^t+X^{(24)}(X^{(24)})^t)(\hat{e}^{(j_1)})^t \right] \right. \nonumber \\
   & &  \left. +\sum_{j_1=1}^{2} \hat{e}^{(j_1)}(-1)^{j_1-1}\left[(\vec{s}^{(2)}(\vec{y}^{(2)})^t+T^{(24)}(X^{(24)})^t) (\hat{e}^{(j_1)})^t   \right. \right. \nonumber \\
   & &+\left. \left. \hat{e}^{(j_1)} (\vec{y}^{(2)}(\vec{s}^{(2)})^t+X^{(24)}(T^{(24)})^t) (\hat{e}^{(j_1)})^t\right] \right.  \nonumber \\
   & &  \left.+\sum_{j_1=1}^{2}\sum_{j_2=1}^{2}\frac{1}{4}\left[\hat{e}^{(j_1j_2)}(\vec{s}^{(3)}(\vec{s}^{(3)})^t
   +T^{(34)}(T^{(34)})^t) (\hat{e}^{(j_1j_2)})^t  \right. \right.   \nonumber \\
   & &+ \left. \left.\hat{e}^{(j_1j_2)} (\vec{y}^{(3)}(\vec{y}^{(3)})^t+X^{(34)}(X^{(34)})^t) (\hat{e}^{(j_1j_2)})^t \right. \right.   \nonumber \\
   & &\left. \left. +\hat{e}^{(j_1j_2)}(\vec{y}^{(3,j_1)}(\vec{y}^{(3,j_1)})^t+X^{(34,j_1)}(X^{(34,j_1)})^t) (\hat{e}^{(j_1j_2)})^t \right. \right.\nonumber \\
   & & \left. \left.  +\hat{e}^{(j_1j_2)}(\vec{z}^{(3,j_1)}(\vec{z}^{(3,j_1)})^t+W^{(34,j_1)}(W^{(34,j_1)})^t) (\hat{e}^{(j_1j_2)})^t \right]    \right.\nonumber \\
   & & \left. +\sum_{j_1=1}^{2}\sum_{j_2=1}^{2}\frac{1}{4}(-1)^{j_1-1}\left[\hat{e}^{(j_1j_2)}(\vec{s}^{(3)}(\vec{y}^{(3)})^t
   +T^{(34)}(X^{(34)})^t) (\hat{e}^{(j_1j_2)})^t  \right. \right.   \nonumber \\
   & &+ \left. \left.\hat{e}^{(j_1j_2)} (\vec{y}^{(3)}(\vec{s}^{(3)})^t+X^{(34)}(T^{(34)})^t) (\hat{e}^{(j_1j_2)})^t \right. \right.   \nonumber \\
   & & \left. \left. +\hat{e}^{(j_1j_2)}(\vec{y}^{(3,j_1)}(\vec{z}^{(3,j_1)})^t+X^{(34,j_1)}(W^{(34,j_1)})^t) (\hat{e}^{(j_1j_2)})^t \right. \right.\nonumber \\
   & & \left. \left.  +\hat{e}^{(j_1j_2)}(\vec{z}^{(3,j_1)}(\vec{y}^{(3,j_1)})^t+W^{(34,j_1)}(X^{(34,j_1)})^t) (\hat{e}^{(j_1j_2)})^t \right]    \right.\nonumber \\
   & & + \left. \sum_{j_1=1}^{2}\sum_{j_2=1}^{2}\frac{1}{4}(-1)^{j_2-1}\left[\hat{e}^{(j_1j_2)}(\vec{s}^{(3)}(\vec{y}^{(3,j_1)})^t
   +T^{(34)}(X^{(34,j_1)})^t) (\hat{e}^{(j_1j_2)})^t  \right. \right.   \nonumber \\
   & &+ \left. \left.\hat{e}^{(j_1j_2)} (\vec{y}^{(3,j_1)}(\vec{s}^{(3)})^t+X^{(34,j_1)}(T^{(34)})^t) (\hat{e}^{(j_1j_2)})^t \right. \right.   \nonumber \\
   & &  + \left. \left.\hat{e}^{(j_1j_2)}(\vec{y}^{(3)}(\vec{z}^{(3,j_1)})^t+X^{(34)}(W^{(34,j_1)})^t) (\hat{e}^{(j_1j_2)})^t \right. \right.\nonumber \\
   & &+ \left. \left.  \hat{e}^{(j_1j_2)}(\vec{z}^{(3,j_1)}(\vec{y}^{(3)})^t+W^{(34,j_1)}(X^{(34)})^t) (\hat{e}^{(j_1j_2)})^t \right]  \right.\nonumber \\
   & &  + \left. \sum_{j_1=1}^{2}\sum_{j_2=1}^{2}\frac{1}{4}(-1)^{j_1-1}(-1)^{j_2-1}\left[\hat{e}^{(j_1j_2)}\left(\vec{s}^{(3)}(\vec{z}^{(3,j_1)})^t
    \right. \right. \right.  \nonumber \\
    & &+ \left. \left. \left. T^{(34)}(W^{(34,j_1)})^t\right) (\hat{e}^{(j_1j_2)})^t  \right. \right.   \nonumber \\
   & &+ \left. \left.\hat{e}^{(j_1j_2)} (\vec{z}^{(3,j_1)}(\vec{s}^{(3)})^t+W^{(34,j_1)}(T^{(34)})^t) (\hat{e}^{(j_1j_2)})^t \right. \right.   \nonumber \\
   & & \left. \left. +\hat{e}^{(j_1j_2)}(\vec{y}^{(3)}(\vec{y}^{(3,j_1)})^t+X^{(34)}(X^{(34,j_1)})^t) (\hat{e}^{(j_1j_2)})^t \right. \right.\nonumber \\
   & & \left. \left.  +\hat{e}^{(j_1j_2)}(\vec{y}^{(3,j_1)}(\vec{y}^{(3)})^t+X^{(34,j_1)}(X^{(34)})^t) (\hat{e}^{(j_1j_2)})^t \right]  \right\} \nonumber \\
\end{eqnarray}
or,

\begin{eqnarray}\label{e50}
  ||\mathcal{C}\times_1 A_1 \times_2 A_2^{(j_1)}\times_3 A_3^{(j_1j_2)}||^2 &=& \frac{1}{2^4}\left\{1+||\vec{s}^{(4)}||^2 + \hat{e} \;\big{[} \vec{s}^{(1)}(\vec{s}^{(1)})^t  +\hat{e} T^{(14)}(T^{(14)})^t\big{]} \hat{e}^t  \right.  \nonumber \\
  & &\left. +\sum_{j_1=1}^{2}\hat{e}^{(j_1)}\frac{1}{2}\left[\vec{s}^{(2)}(\vec{s}^{(2)})^t+T^{(24)}(T^{(24)})^t +\vec{y}^{(2)}(\vec{y}^{(2)})^t+X^{(24)}(X^{(24)})^t \right. \right. \nonumber \\
 & &+\left. \left.  (-1)^{j_1-1}\left(\vec{s}^{(2)}(\vec{y}^{(2)})^t+T^{(24)}(X^{(24)})^t  \right. \right.  \right. \nonumber \\
    & &+\left. \left. \left.  \vec{y}^{(2)}(\vec{s}^{(2)})^t+X^{(24)}(T^{(24)})^t \right) \right](\hat{e}^{(j_1)})^t \right.   \nonumber \\
   & &  \left.+\sum_{j_1=1}^{2}\sum_{j_2=1}^{2}\hat{e}^{(j_1j_2)}\frac{1}{4}\left[\vec{s}^{(3)}(\vec{s}^{(3)})^t
   +T^{(34)}(T^{(34)})^t \right. \right.   \nonumber \\
   & &+ \left. \left. \vec{y}^{(3)}(\vec{y}^{(3)})^t+X^{(34)}(X^{(34)})^t  \right. \right.   \nonumber \\
   & &+ \left. \left. \vec{y}^{(3,j_1)}(\vec{y}^{(3,j_1)})^t+X^{(34,j_1)}(X^{(34,j_1)})^t  \right. \right.\nonumber \\
   & & +\left. \left. \vec{z}^{(3,j_1)}(\vec{z}^{(3,j_1)})^t+W^{(34,j_1)}(W^{(34,j_1)})^t \right.    \right.\nonumber \\
   & &+ \left. \left. (-1)^{j_1-1}\left(\vec{s}^{(3)}(\vec{y}^{(3)})^t
   +T^{(34)}(X^{(34)})^t  \right. \right. \right.   \nonumber \\
   & &+ \left. \left. \left. \vec{y}^{(3)}(\vec{s}^{(3)})^t+X^{(34)}(T^{(34)})^t  \right. \right. \right.  \nonumber \\
   & &+ \left. \left. \left. \vec{y}^{(3,j_1)}(\vec{z}^{(3,j_1)})^t+X^{(34,j_1)}(W^{(34,j_1)})^t    \right.  \right. \right.\nonumber \\
   & &+ \left. \left. \left.  \vec{z}^{(3,j_1)}(\vec{y}^{(3,j_1)})^t+W^{(34,j_1)}(X^{(34,j_1)})^t  \right)  \right. \right.  \nonumber \\
   & & + \left. \left.  (-1)^{j_2-1}\left(\vec{s}^{(3)}(\vec{y}^{(3,j_1)})^t
   +T^{(34)}(X^{(34,j_1)})^t  \right. \right. \right.   \nonumber \\
   & &+\left. \left. \left. \vec{y}^{(3,j_1)}(\vec{s}^{(3)})^t+X^{(34,j_1)}(T^{(34)})^t \right. \right. \right.   \nonumber \\
   & &  +\left. \left. \left. \vec{y}^{(3)}(\vec{z}^{(3,j_1)})^t+X^{(34)}(W^{(34,j_1)})^t \right. \right. \right.\nonumber \\
   & &+\left. \left. \left. \vec{z}^{(3,j_1)}(\vec{y}^{(3)})^t+W^{(34,j_1)}(X^{(34)})^t  \right) \right. \right.\nonumber \\
   & &  + \left. \left. (-1)^{j_1-1}(-1)^{j_2-1}\left(\vec{s}^{(3)}(\vec{z}^{(3,j_1)})^t
   +  T^{(34)}(W^{(34,j_1)})^t   \right. \right. \right.   \nonumber \\
   & &+ \left. \left. \left. \vec{z}^{(3,j_1)}(\vec{s}^{(3)})^t+W^{(34,j_1)}(T^{(34)})^t \right. \right. \right.   \nonumber \\
   & &+ \left. \left. \left.  \vec{y}^{(3)}(\vec{y}^{(3,j_1)})^t+X^{(34)}(X^{(34,j_1)})^t \right. \right. \right.\nonumber \\
   & &+ \left. \left. \left.  \vec{y}^{(3,j_1)}(\vec{y}^{(3)})^t+X^{(34,j_1)}(X^{(34)})^t  \right) \right]  (\hat{e}^{(j_1j_2)})^t \right\} \nonumber \\
\end{eqnarray}
where $ \vec{y}^{(2)}_{i_2}=\sum_{i_1=1}^{3} \hat{e}_{i_1} t_{i_1i_2}^{(12)},\; X_{i_2i_4}^{(24)}=\sum_{i_1}^{3} \hat{e}_{i_1} t_{i_1i_2 i_4}^{(124)},\;\; \vec{y}^{(3)}_{i_3}=\sum_{i_1=1}^{3} \hat{e}_{i_1} t_{i_1i_3}^{(13)},$ \\
$ X_{i_3i_4}^{(34)}=\sum_{i_1}^{3} \hat{e}_{i_1} t_{i_1i_3 i_4}^{(134)},$
and
$\;\;  \vec{y}^{(3,j_1)}_{i_3}=\sum_{i_2=1}^{3} \hat{e}_{i_2}^{(j_1)} t_{i_2i_3}^{(23)},\; X_{i_3i_4}^{(34,j_1)}=\sum_{i_2}^{3} \hat{e}_{i_2}^{(j_1)} t_{i_2i_3 i_4}^{(234)},$\\
$\vec{z}^{(3,j_1)}_{i_3}=\sum_{i_1=1}^{3} \sum_{i_2=1}^{3} \hat{e}_{i_1} \hat{e}_{i_2}^{(j_1)} t_{i_1i_2i_3}^{(123)},\; \; W_{i_3i_4}^{(34,j_1)}=\sum_{i_1}^{3} \sum_{i_2=1}^{3}\hat{e}_{i_1}\hat{e}_{i_2}^{(j_1)} t_{i_1i_2i_3 i_4}^{(1234)}.$\\

Let us replace the square bracketed expressions in Eq.(\ref{e50}), by the  $(3\times 3)$ real symmetric  matrices  as

\begin{eqnarray*}
G  &=& \vec{s}^{(1)}(\vec{s}^{(1)})^t+ T^{(14)}(T^{(14)})^t,  \nonumber \\
  G^{(j_1)} &=& \frac{1}{2}[\vec{s}^{(2)}(\vec{s}^{(2)})^t+\vec{y}^{(2)}(\vec{y}^{(2)})^t+ T^{(24)}(T^{(24)})^t+X^{(24)}(X^{(24)})^t   \nonumber  \\
   & & +(-1)^{j_1-1}(\vec{s}^{(2)}(\vec{y}^{(2)})^t+\vec{y}^{(2)}(\vec{s}^{(2)})^t+T^{(24)}(X^{(24)})^t+X^{(24)}(T^{(24)})^t)], \; j_1=1,2 \nonumber  \\
  G^{(j_1j_2)} &=& \frac{1}{4}[\vec{s}^{(3)}(\vec{s}^{(3)})^t+ T^{(34)}(T^{(34)})^t+\vec{y}^{(3)}(\vec{y}^{(3)})^t
  +X^{(34)}(X^{(34)})^t+ \vec{y}^{(3,j_1)}(\vec{y}^{(3,j_1)})^t +X^{(34,j_1)}(X^{(34,j_1)})^t    \nonumber \\
   & & + \vec{z}^{(3,j_1)}(\vec{z}^{(3,j_1)})^t +W^{(34,j_1)}(W^{(34,j_1)})^t+(-1)^{j_1-1} [\vec{s}^{(3)} (\vec{y}^{(3)})^t + T^{(34)}(X^{(34)})^t \nonumber  \\
   & & +\vec{y}^{(3)}(\vec{s}^{(3)})^t+X^{(34)}(T^{(34)})^t +\vec{y}^{(3,j_1)}(\vec{z}^{(3,j_1)})^t +X^{(34,j_1)}(W^{(34,j_1)})^t+\vec{z}^{(3,j_1)}(\vec{y}^{(3,j_1)})^t\nonumber  \\
   & &+ W^{(34,j_1)}(X^{(34,j_1)})^t]+(-1)^{j_2-1} [\vec{s}^{(3)} (\vec{y}^{(3,j_1)})^t + T^{(34)}(X^{(34,j_1)})^t \nonumber  \\
   & & +\vec{y}^{(3,j_1)}(\vec{s}^{(3)})^t+X^{(34,j_1)}(T^{(34)})^t +\vec{y}^{(3)}(\vec{z}^{(3,j_1)})^t +X^{(34)}(W^{(34,j_1)})^t+\vec{z}^{(3,j_1)}(\vec{y}^{(3)})^t  \nonumber  \\
   & &+ W^{(34,j_1)}(X^{(34)})^t]+(-1)^{j_1-1}(-1)^{j_2-1}[\vec{s}^{(3)}(\vec{z}^{(3,j_1)})^t+T^{(34)}(W^{(34,j_1)})^t  +\vec{z}^{(3,j_1)}(\vec{s}^{(3)})^t   \nonumber  \\
   & &+W^{(34,j_1)}(T^{(34)})^t+\vec{y}^{(3)} (\vec{y}^{(3,j_1)})^t + X^{(34)}(X^{(34,j_1)})^t+\vec{y}^{(3,j_1)}(\vec{y}^{(3)})^t+X^{(34,j_1)}(X^{(34)})^t],
 \end{eqnarray*}
Thus, we get

\begin{eqnarray}\label{e51}
  ||\mathcal{C}\times_1 A_1 \times_2 A_2^{(j_1)}\times_3 A_3^{(j_1j_2)}||^2 &=& \frac{1}{2^4}\left\{1+||\vec{s}^{(4)}||^2 + \hat{e} G \hat{e}^t + \sum_{j_1=1}^{2}\hat{e}^{(j_1)} G^{(j_1)} (\hat{e}^{(j_1)})^t \right. \nonumber \\
   & & +\left. \sum_{j_1=1}^{2} \sum_{j_2=1}^{2}\hat{e}^{(j_1j_2)} G^{(j_1j_2)} (\hat{e}^{(j_1j_2)})^t \right\}.
\end{eqnarray}
In Eq.(\ref{e51}) the three last terms  {depend} on matrices $A_1,\; A_2^{(j_1)},$  {and} $A_3^{(j_1j_2)}$ while others
are determined by the state $\rho_{1234}.$ Therefore, to maximize
$||\mathcal{C}\times_1 A_1 \times_2 A_2^{(j_1)}\times_3 A_3^{(j_1j_2)}||^2$ we take $\hat{e},\; \hat{e}^{(j_1)},\; \hat{e}^{(j_1j_2)}$
to be the eigenvectors of $G, \; G^{(j_1)},\; G^{(j_1j_2)}$ corresponding to its largest eigenvalues
$\eta_{max},\;\eta_{max}^{(j_1)}, \; \eta_{max}^{(j_1j_2)}$ respectively,  so that

\begin{equation}\label{e52}
\max_{A_1,A_2^{(j_1)},A_3^{(j_1j_2)}}||\mathcal{C}\times_1 A_1 \times_2 A_2^{(j_1)}\times_3 A_2^{(j_1j_2)}||^2 = \frac{1}{2^4}\left\{1+||\vec{s}^{(4)}||^2 + \eta_{max}+\sum_{j_1=1}^{2}\eta_{max}^{(j_1)}+\sum_{j_1=1}^{2}\sum_{j_2=1}^{2}\eta_{max}^{(j_1j_2)} \right\}.
\end{equation}
Finally, Eq.(\ref{e52}), Eq.(\ref{e31}) and  Eq.(\ref{e10}) together imply

\begin{eqnarray}\label{e53}
   GD_{A_1;A_2;A_3;A_4}(\rho_{1234}) &=& \frac{1}{2^4}\left\{||\vec{s}^{(1)}||^2+||\vec{s}^{(2)}||^2+||\vec{s}^{(3)}||^2+\sum_{1\leq k< l \leq 4}||T^{(kl)}||^2+ \sum_{1\leq k< l< q \leq 4}||T^{(klq)}||^2 \right. \nonumber \\
   & & \left.+||\mathcal{T}||^2-\eta_{max}-\sum_{j_1=1}^{2}\eta_{max}^{(j_1)}-\sum_{j_1=1}^{2} \sum_{j_2=1}^{2}\eta_{max}^{(j_1j_2)}\right\},
 \end{eqnarray}
where $\eta_{max},\;\eta_{max}^{(j_1)}, \; \eta_{max}^{(j_1j_2)},\; j_1,j_2=1,2$ are the largest eigenvalues of matrices
$G, \; G^{(j_1)},\; G^{(j_1j_2)}$ respectively, thus completing the proof.\\
In general, we can prove for $N$-qubit quantum state\\
\textbf{Theorem 4}. Let $\rho_{1\cdots N}$ be  {an} $N$-qubit state defined by Eq.(\ref{e18}), then
\begin{eqnarray}\label{e54}
GD_{A_1;\cdots;A_N}(\rho_{1\cdots N}) & = & \frac{1}{2^N} \left[\sum_{k=1}^{N-1} ||\vec{s}^{(k)}||^2+\sum_{\{k_1,k_2\}} ||T^{\{k_1,k_2\}}||^2+\cdots+\sum_{\{k_1,\cdots,k_M\}}  ||T^{\{k_1,k_2,\cdots,k_M\}}||^2 \right. \nonumber \\
& &+\left. \cdots+ ||\mathcal{T}||^2-\eta_{max}-\sum_{m=3}^{N} \sum_{j_1\cdots j_m=1}^{2}\eta_{max}^{(j_1\cdots j_{m-2})} \right].\nonumber \\
\end{eqnarray}
Here $\eta_{max},\;\;\eta_{max}^{(j_1j_2\cdots j_{m-2})},\; 3\leq m \leq N$  are the largest eigenvalues, corresponding
to $\hat{e},\;\hat{e}^{(j_1\cdots j_{m-2})}$ eigenvectors, of the matrix
$G,\; \;  G^{(j_1j_2\cdots j_{m-2})},\; \; j_1,\cdots j_{m-2}=1,2$ which is a $3 \times 3$ real symmetric matrices, defined as

\begin{eqnarray} \label{e55}
G  &=& \vec{s}^{(1)}(\vec{s}^{(1)})^t+ T^{(1m)}(T^{(1m)})^t,  \nonumber \\
G^{(j_1j_2\cdots j_{m-2})} &=& \frac{1}{2^{m-2}}\left\{ \vec{s}^{(m-1)}(\vec{s}^{(m-1)})^t+ T^{(m-1m)}(T^{(m-1m)})^t+\vec{y}^{(m-1)}(\vec{y}^{(m-1)})^t+X^{(m-1m)}(X^{(m-1m)})^t\right. \nonumber \\
& & \left.+(-1)^{j_1-1}\left[\vec{s}^{(m)}(\vec{y}^{(m)})^t+ T^{(mN)}(X^{(mN)})^t+\vec{y}^{(m)}(\vec{s}^{(m)})^t+ X^{(mN)}(T^{(mN)})^t  \right] \right.\nonumber \\
& & +\left. \sum_{k=1}^{m-3}\left[\vec{y}^{(m-1,j_k)}(\vec{y}^{(m-1,j_k)})^t +X^{(m-1m,j_k)}(X^{(m-1m,j_k)})^t \right. \right.  \nonumber \\
& &\left. \left. +\vec{z}^{(m-1,j_k)}(\vec{z}^{(m-1,j_k)})^t +W^{(m-1m,j_k)}(W^{(m-1m,j_k)})^t\right.\right.\nonumber \\
& &\left. \left. +(-1)^{j_{k+1}-1} \left[\vec{s}^{(m-1)}(\vec{y}^{(m-1,j_k)})^t+ T^{(m-1m)}(X^{(m-1m,j_k)})^t+\vec{y}^{(m-1,j_k)}(\vec{s}^{(m-1)})^t
    \right. \right. \right. \nonumber\\
& & \left. \left. \left.+ X^{(m-1m,j_k)}(T^{(m-1m)})^t+\vec{y}^{(m-1)}(\vec{z}^{(m-1,j_k)})^t+ X^{(m-1m)}(W^{(m-1m,j_k)})^t
\right. \right.\right. \nonumber\\
& &  \left.\left.\left.  +\vec{z}^{(m-1,j_k)}(\vec{y}^{(m-1)})^t+ W^{(m-1m,j_k)}(X^{(m-1m)})^t+(-1)^{j_1-1}\left[\vec{s}^{(m-1)}(\vec{z}^{(m-1,j_k)})^t \right.\right. \right.\right. \nonumber\\
& & \left. \left. \left.\left. + T^{(m-1m)}(W^{(m-1m,j_k)})^t+\vec{z}^{(m-1,j_k)}(\vec{s}^{(m-1)})^t+ W^{(m-1m,j_k)}(T^{(m-1m)})^t\right. \right.\right.\right. \nonumber\\
& & \left. \left. \left.\left.+\vec{y}^{(m-1)}(\vec{y}^{(m-1,j_k)})^t+ X^{(m-1m)}(X^{(m-1m,j_k)})^t+\vec{y}^{(m-1,j_k)}(\vec{y}^{(m-1)})^t\right. \right.\right.\right. \nonumber\\
& & \left.\left.\left. \left. + X^{(m-1m,j_k)}(X^{(m-1m)})^t \right] \right]\right]\right. \nonumber\\
& &\left. +\sum_{M=2}^{m-3} \sum_{\{k_1,k_2,\cdots,k_M\}}  \left[\vec{y}^{(m-1,\{k_1,k_2,\cdots,k_M\})}(\vec{y}^{(m-1,\{k_1,k_2,\cdots,k_M\})})^t\right.\right.\nonumber \\
& & \left. \left. + X^{(m-1m,\{k_1,k_2,\cdots,k_M\})}(X^{(m-1m,\{k_1,k_2,\cdots,k_M\})})^t+\vec{z}^{(m-1,\{k_1,k_2,\cdots,k_M\})}(\vec{z}^{(m-1,\{k_1,k_2,\cdots,k_M\})})^t \right. \right. \nonumber \\
& &+\left. \left. W^{(m-1m,\{k_1,k_2,\cdots,k_M\})}(W^{(m-1m,\{k_1,k_2,\cdots,k_M\})})^t \right.\right. \nonumber\\
& &\left. \left. +(-1)^{j_{k_1+1}-1} (-1)^{j_{k_2+1}-1}\cdots (-1)^{j_{k_M+1}-1}\left[\vec{s}^{(m-1)}(\vec{y}^{(m-1,\{j_{k_1},j_{k_2},\cdots,j_{k_M}\})})^t\right. \right.\right. \nonumber \\
& & \left. \left. \left.+ T^{(m-1m)}(X^{(m-1m,\{j_{k_1},j_{k_2},\cdots,j_{k_M}\})})^t+\vec{y}^{(m-1,\{j_{k_1},j_{k_2},\cdots,j_{k_M}\})}(\vec{s}^{(m-1)})^t \right. \right.\right. \nonumber\\
& & \left. \left. \left.+ X^{(m-1m,\{j_{k_1},j_{k_2},\cdots,j_{k_M}\})}(T^{(m-1m)})^t+\vec{y}^{(m-1)}(\vec{z}^{(m-1,\{j_{k_1},j_{k_2},\cdots,j_{k_M}\})})^t \right. \right.\right. \nonumber\\
& &\left. \left.\left. + X^{(m-1m)}(W^{(m-1m,\{j_{k_1},j_{k_2},\cdots,j_{k_M}\})})^t+\vec{z}^{(m-1,\{j_{k_1},j_{k_2},\cdots,j_{k_M}\})}(\vec{y}^{(m-1)})^t \right. \right.\right. \nonumber\\
& &\left. \left.\left. + W^{(m-1m,\{j_{k_1},j_{k_2},\cdots,j_{k_M}\})}(X^{(m-1m)})^t +(-1)^{j_1-1}\left[\vec{s}^{(m-1)}(\vec{z}^{(m-1,\{j_{k_1},j_{k_2},\cdots,j_{k_M}\})})^t \right. \right.\right.\right. \nonumber\\
& &\left. \left.\left. \left.+ T^{(m-1m)}(W^{(m-1m,\{j_{k_1},j_{k_2},\cdots,j_{k_M}\})})^t+\vec{z}^{(m-1,\{j_{k_1},j_{k_2},\cdots,j_{k_M}\})}(\vec{s}^{(m-1)})^t \right.\right.\right.\right. \nonumber\\
& &\left. \left.\left. \left. + W^{(m-1m,\{j_{k_1},j_{k_2},\cdots,j_{k_M}\})}(T^{(m-1m)})^t+\vec{y}^{(m-1)}(\vec{y}^{(m-1,\{j_{k_1},j_{k_2},\cdots,j_{k_M}\})})^t \right.\right.\right.\right. \nonumber\\
& &\left. \left.\left.\left. +X^{(m-1m)}(X^{(m-1m,\{j_{k_1},j_{k_2},\cdots,j_{k_M}\})})^t+\vec{y}^{(m-1,\{j_{k_1},j_{k_2},\cdots,j_{k_M}\})}(\vec{y}^{(m-1)})^t
 \right.\right.\right.\right. \nonumber\\
& &\left. \left.\left.\left. + X^{(m-1m,\{j_{k_1},j_{k_2},\cdots,j_{k_M}\})}(X^{(m-1m)})^t \right]\right]\right]+ F \right\}.\nonumber\\
\end{eqnarray}
The matrix $F$ is defined as,
\begin{eqnarray*}
F & = &\sum_{k=1}^{m-3}\sum_{M=2}^{m-3} \sum_{\{k_1,\cdots,k_M\}} (-1)^{j_{k+1}-1}(-1)^{j_{k_1+1}-1} (-1)^{j_{k_2+1}-1}\cdots(-1)^{j_{k_M+1}-1}\nonumber\\
& & \times \left[\vec{y}^{(m-1,j_k)}(\vec{y}^{(m-1,\{j_{k_1},j_{k_2},\cdots,j_{k_M}\})})^t+ X^{(m-1m,j_k)}(X^{(m-1m,\{j_{k_1},j_{k_2},\cdots,j_{k_M}\})})^t\right.  \nonumber\\
& &\left.+\vec{y}^{(m-1,\{j_{k_1},j_{k_2},\cdots,j_{k_M}\})}(\vec{y}^{(m-1,j_k)})^t + X^{(m-1m,\{j_{k_1},j_{k_2},\cdots,j_{k_M}\})}(X^{(m-1m,j_k)})^t \right.  \nonumber\\
& & \left.+(-1)^{j_1-1} \left[\vec{y}^{(m-1,j_k)}(\vec{z}^{(m-1,\{j_{k_1},j_{k_2},\cdots,j_{k_M}\})})^t + X^{(m-1m,j_k)}(W^{(m-1m,\{j_{k_1},j_{k_2},\cdots,j_{k_M}\})})^t \right.\right.  \nonumber\\
& & \left.\left.+\vec{z}^{(m-1,\{j_{k_1},j_{k_2},\cdots,j_{k_M}\})}(\vec{y}^{(m-1,j_k)})^t+ W^{(m-1m,\{j_{k_1},j_{k_2},\cdots,j_{k_M}\})}(X^{(m-1m,j_k)})^t \right]\right]  \nonumber\\
& & +\sum_{M=2}^{m-3} \sum_{L=2}^{m-3}\sum_{\{k_1,k_2,\cdots,k_M\}}\sum_{\{q_1,q_2,\cdots,q_L\}\neq \{k_1,k_2,\cdots,k_M\}} (-1)^{j_{q_1+1}-1}\cdots (-1)^{j_{q_L+1}-1}(-1)^{j_{k_1+1}-1} \cdots \nonumber\\
& &\times (-1)^{j_{k_M+1}-1} \left[\vec{y}^{(m-1,\{j_{q_1},j_{q_2},\cdots,j_{q_L}\})}(\vec{y}^{(m-1,\{j_{k_1},j_{k_2},\cdots,j_{k_M}\})})^t \right.\nonumber\\
& & \left.+X^{(m-1m,\{j_{q_1},j_{q_2},\cdots,j_{q_L}\})}(X^{(m-1m,\{j_{k_1},j_{k_2},\cdots,j_{k_M}\})})^t \right.\nonumber\\
& & \left. +\vec{y}^{(m-1,\{j_{k_1},j_{k_2},\cdots,j_{k_M}\})}(\vec{y}^{(m-1,\{j_{q_1},j_{q_2},\cdots,j_{q_L}\})})^t\right. \nonumber\\
& & \left.+X^{(m-1m,\{j_{k_1},j_{k_2},\cdots,j_{k_M}\})}(X^{(m-1m,\{j_{q_1},j_{q_2},\cdots,j_{q_L}\})})^t \right. \nonumber\\
& &\left. + \vec{z}^{(m-1,\{j_{q_1},j_{q_2},\cdots,j_{q_L}\})}(\vec{z}^{(m-1,\{j_{k_1},j_{k_2},\cdots,j_{k_M}\})})^t\right. \nonumber\\
& &\left.+W^{(m-1m,\{j_{q_1},j_{q_2},\cdots,j_{q_L}\})}(W^{(m-1m,\{j_{k_1},j_{k_2},\cdots,j_{k_M}\})})^t\right. \nonumber\\
& &\left.+\vec{z}^{(m-1,\{j_{k_1},j_{k_2},\cdots,j_{k_M}\})}(\vec{z}^{(m-1,\{j_{q_1},j_{q_2},\cdots,j_{q_L}\})})^t\right.\nonumber\\
& &\left.+W^{(m-1m,\{j_{k_1},j_{k_2},\cdots,j_{k_M}\})}(W^{(m-1m,\{j_{q_1},j_{q_2},\cdots,j_{q_L}\})})^t\right.\nonumber\\
& & \left.+ (-1)^{j_1-1}\left[\vec{y}^{(m-1,\{j_{q_1},j_{q_2},\cdots,j_{q_L}\})}(\vec{z}^{(m-1,\{j_{k_1},j_{k_2},\cdots,j_{k_M}\})})^t \right.\right.\nonumber\\
& & \left.\left.+ X^{(m-1m,\{j_{q_1},j_{q_2},\cdots,j_{q_L}\})}(W^{(m-1m,\{j_{k_1},j_{k_2},\cdots,j_{k_M}\})})^t \right.\right.\nonumber\\
& & \left.\left.+\vec{z}^{(m-1,\{j_{k_1},j_{k_2},\cdots,j_{k_M}\})}(\vec{y}^{(m-1,\{j_{q_1},j_{q_2},\cdots,j_{q_L}\})})^t \right.\right.\nonumber\\
& & \left.\left.+ W^{(m-1m,\{j_{k_1},j_{k_2},\cdots,j_{k_M}\})}(X^{(m-1m,\{j_{q_1},j_{q_2},\cdots,j_{q_L}\})})^t\right]\right]\nonumber\\
\end{eqnarray*}
where $\vec{y}^{(m-1)}_{i_{m-1}}=\sum_{i_1=1}^{3} \hat{e}_{i_1} t_{i_1i_{m-1}},\; X_{i_{m-1}i_m}^{(m-1m)}=\sum_{i_1}^{3} \hat{e}_{i_1} t_{i_1i_{m-1} i_m},\; m=3,\cdots, N,$\\

\noindent and for $1\leq k \leq m-3,$\\

$\vec{y}^{(m-1,j_k)}_{i_{m-1}}=\sum_{i_{k+1}}^{3} \hat{e}_{i_{k+1}}^{(j_1\cdots j_k)} t_{i_{k+1}i_{m-1}},$\\

$X_{i_{m-1}i_m}^{(m-1m,j_k)}=\sum_{i_{k+1}}^{3} \hat{e}_{i_{k+1}}^{(j_1\cdots j_k)} t_{i_{k+1}i_{m-1}i_{m}},$\\

$\vec{z}^{(m-1,j_k)}_{i_{m-1}}=\sum_{i_1i_{k+1}}^{3} \hat{e}_{i_1}\hat{e}_{i_{k+1}}^{(j_1\cdots j_k)} t_{i_1i_{k+1}i_{m-1}},$\\

$W_{i_{m-1}i_m}^{(m-1m,j_k)}=\sum_{i_1i_{k+1}}^{3} \hat{e}_{i_1}\hat{e}_{i_{k+1}}^{(j_1\cdots j_k)} t_{i_1i_{k+1}i_{m-1}i_{m}}.$\\

\noindent For every subset $\{k_1,k_2,\cdots,k_M\}$ of $\{1,2,3,\cdots, m-3\},$ and $2\leq M \leq m-3.$\\

$\vec{y}^{(m-1,\{k_1,k_2,\cdots,k_M\})}_{i_{m-1}}=\sum_{i_{k_1+1}i_{k_2+1}\cdots i_{k_M+1}}^{3}  \hat{e}_{i_{k_1+1}}^{(j_1\cdots j_{k_1})} \hat{e}_{i_{k_2+1}}^{(j_1\cdots j_{k_2})}\cdots \hat{e}_{i_{k_M+1}}^{(j_1\cdots j_{k_M})}t_{i_{k_1+1}i_{k_2+1}\cdots i_{k_M+1}i_{m-1}},$\\

$X_{i_{m-1}i_m}^{(m-1m,\{k_1,k_2,\cdots,k_M\})}=\sum_{i_{k_1+1}i_{k_2+1}\cdots i_{k_M+1}}^{3}  \hat{e}_{i_{k_1+1}}^{(j_1\cdots j_{k_1})} \hat{e}_{i_{k_2+1}}^{(j_1\cdots j_{k_2})}\cdots \hat{e}_{i_{k_M+1}}^{(j_1\cdots j_{k_M})}t_{i_{k_1+1}i_{k_2+1}\cdots i_{k_M+1}i_{m-1}i_m},$\\

$\vec{z}^{(m-1,\{k_1,k_2,\cdots,k_M\})}_{i_{m-1}}=\sum_{i_1i_{k_1+1}i_{k_2+1}\cdots i_{k_M+1}}^{3}  \hat{e}_{i_1} \hat{e}_{i_{k_1+1}}^{(j_1\cdots j_{k_1})} \hat{e}_{i_{k_2+1}}^{(j_1\cdots j_{k_2})}\cdots \hat{e}_{i_{k_M+1}}^{(j_1\cdots j_{k_M})}t_{i_1i_{k_1+1}i_{k_2+1}\cdots i_{k_M+1}i_{m-1}},$\\

$W_{i_{m-1}i_m}^{(m-1m,\{k_1,k_2,\cdots,k_M\})}=\sum_{i_1i_{k_1+1}i_{k_2+1}\cdots i_{k_M+1}}^{3}  \hat{e}_{i_1} \hat{e}_{i_{k_1+1}}^{(j_1\cdots j_{k_1})} \hat{e}_{i_{k_2+1}}^{(j_1\cdots j_{k_2})}\cdots \hat{e}_{i_{k_M+1}}^{(j_1\cdots j_{k_M})}t_{i_1i_{k_1+1}i_{k_2+1}\cdots i_{k_M+1}i_{m-1}i_m}.$\\

\textbf{Proof:}
Following the proof of theorem 3,  {Eq.} (\ref{e40}) becomes

\begin{equation}\label{e56}
  a_{2i_{1}}= - a_{1i_{1}}, \;\;\;
  a_{2i_{m-1}}^{(j_1\cdots j_{m-2})}= - a_{1i_{m-1}}^{(j_1\cdots j_{m-2})},  \; j_{m-2}=1,2; \;i_{m-1}=2,3,4; \; m=3,4\cdots N.
\end{equation}

We now proceed to construct the $2\times 4$ matrices  $A_1,\; A_{m-1}^{(j_1j_2\cdots j_{m-2})},\; m=3,4,\cdots N$ defined
via Eq.(\ref{e12}). We will use Eq.(\ref{e56}). The row vectors of $A_1,\; A_{m-1}^{(j_1j_2\cdots j_{m-2})}$ are
$$\vec{a}_{j_1}=(a_{j_11},a_{j_12},a_{j_13},a_{j_14}), \; j_1=1,2.$$
$$\vec{a}_{j_{m-1}}^{(j_1\cdots j_{m-2})}=(a_{j_{m-1}1}^{(j_1\cdots j_{m-2})},a_{j_{m-1}2}^{(j_1\cdots j_{m-2})},a_{j_{m-1}3}^{(j_1\cdots j_{m-2})},a_{j_{m-1}4}^{(j_1\cdots j_{m-2})}),\; j_{m-1}=1,2.$$
Next, we define
\begin{equation}\label{e57}
 \hat{e}_{j_1}=\sqrt{2}(a_{j_12},a_{j_13},a_{j_14}),\;j_1=1,2, \; \; \;
 \hat{e}_{j_{m-1}}^{(j_1\cdots j_{m-2})}=\sqrt{2}(a_{j_{m-1}2}^{(j_1\cdots j_{m-2})},a_{j_{m-1}3}^{(j_1\cdots j_{m-2})},a_{j_{m-1}4}^{(j_1\cdots j_{m-2})}),\;j_{m-1}=1,2,
 \end{equation}
and using Eq.(\ref{e56}), we get
\begin{equation}\label{e58}
 \hat{e}_2=- \hat{e}_1, \;\; \;
  \hat{e}_{2}^{(j_1\cdots j_{m-2})}=- \hat{e}_1^{(j_1\cdots j_{m-2})}.
 \end{equation}
We can prove
\begin{equation}\label{e59}
  ||\hat{e}_{j_1}||^2=1 ,\;\;j_1=1,2, \; \; \;
  ||\hat{e}_{j_{m-1}}^{(j_1\cdots j_{m-2})}||^2=1 ,\;\;j_{m-1}=1,2,
 \end{equation}
using the condition Eq.(\ref{e13}), $||\vec{a}_{j_1}||^2=\sum_{i_1=1}^{4} a_{j_1i_{1}}^2=1,\; ||\vec{a}_{j_{m-1}}^{(j_1\cdots j_{m-2})}||^2=\sum_{i_{m-1}=1}^{4} (a_{j_{m-1}i_{m-1}}^{(j_1\cdots j_{m-2})})^2=1,$
and $a_{j_11}= tr(|j_1\rangle \langle j_1| X^{(1)}_1)=\frac{1}{\sqrt{2}}, \; \; a_{j_{m-1}}^{(j_1\cdots j_{m-2})}=tr(|j_{m-1};j_1\cdots j_{m-2}\rangle \langle j_{m-1};j_1\cdots j_{m-2}| X^{(m-1)}_1)=\frac{1}{\sqrt{2}}.$\\

We can now construct the row vectors of $2\times 4$ matrices $A_1,\; A_{m-1}^{(j_1j_2\cdots j_{m-2})},$
using Eq.(\ref{e57}) and Eq.(\ref{e58}), with defining
$\hat{e}_1=\hat{e},\;\hat{e}_1^{(j_1\cdots j_{m-2})}=\hat{e}^{(j_1\cdots j_{m-2})},\; j_{m-2}=1,2,\; m=3,\cdots N$

\begin{equation}\label{e60}
  \vec{a}_1=\frac{1}{\sqrt{2}} (1, \hat{e}),  \; \; \;
  \vec{a}_2=\frac{1}{\sqrt{2}} (1, -\hat{e}),
  \end{equation}
\begin{equation}\label{e61}
  \vec{a}_1^{(j_1\cdots j_{m-2})}=\frac{1}{\sqrt{2}} (1, \hat{e}^{(j_1\cdots j_{m-2})}),  \; \; \;
  \vec{a}_2^{(j_1\cdots j_{m-2})}=\frac{1}{\sqrt{2}} (1, -\hat{e}^{(j_1\cdots j_{m-2})}),
\end{equation}

The matrix $A_1$ and the conditional matrices $A_{m-1}^{(j_1\cdots j_{m-2})}$ for 1st, 2nd to $m$th ($m=3,\cdots N$)
systems respectively are, in terms of the row vectors defined above,

\begin{displaymath}
A_1=\frac{1}{\sqrt{2}}
\left(\begin{array}{cc}
1 & \hat{e}\\
1 & -\hat{e}\\
\end{array}\right),
\end{displaymath}
and

\begin{displaymath}
A_{m-1}^{(j_1\cdots j_{m-2})}=\frac{1}{\sqrt{2}}
\left(\begin{array}{cc}
1 & \hat{e}^{(j_1\cdots j_{m-2})}\\
1 & -\hat{e}^{(j_1\cdots j_{m-2})}\\
\end{array}\right).
\end{displaymath}

The norm of the tensor $\mathcal{C}$ can be expressed in terms of the norms of the tensors defining $\rho_{12\cdots N}$
by using the equivalence of the definitions of $\rho_{12\cdots N}$ given in Eq.(\ref{e7}) and Eq.(\ref{e18}) as

\begin{eqnarray}\label{e62}
  ||\mathcal{C}||^2 & = & \frac{1}{2^N}\left[1+ \sum_{k_1 \in \mathcal{N}} ||\vec{s}^{(k_1)}||^2+\sum_{\{k_1,k_2\}} ||T^{\{k_1,k_2\}}||^2+\cdots \right.\nonumber \\
 & & \:\left. + \sum_{\{k_1,k_2,\cdots,k_M\}}  ||T^{\{k_1,k_2,\cdots,k_M\}}||^2+\cdots+ + ||\mathcal{T}||^2\right]. \nonumber \\
\end{eqnarray}

In order to get the norm of $\mathcal{C}\times_1 A_1\times_2 A_2^{(j_1)}\times \cdots\times_{N-1} A_{N-1}^{(j_1j_2\cdots j_{N-2})}$
we use its elementwise definition,

\begin{eqnarray}\label{e63}
 (\mathcal{C}\times_1 A_1\times_2 A_2^{(j_1)}\times \cdots\times_{N-1} A_{N-1}^{(j_1j_2\cdots j_{N-2})})_{j_1\cdots j_{N-1} i_N}  &=& \sum_{i_1i_2\cdots i_{N-1}} c_{i_1i_2 \cdots i_N} a_{j_1i_1}a_{j_2 i_2}^{(j_1)}\cdots a_{j_{N-1} i_{N-1}}^{(j_1j_2\cdots j_{N-2})}, \nonumber \\
\end{eqnarray}

Let us put $i_k=0,1,2,3$ instead of $i_k=1,2,3,4,\; k=1,\cdots N,$ the norm can be obtained as,

\begin{eqnarray}\label{e64}
  ||\mathcal{C}\times_1 A_1 \times_2 A_2^{(j_1)}\times_3 \cdots \times_{N-1} A_{N-1}^{(j_1j_2\cdots j_{N-2})}||^2 &=& \sum_{j_1\cdots j_{N-1}}^{2} \sum_{i_N=0}^{3}\left(\sum_{i_1\cdots i_{N-1}} \mathcal{C}_{i_1\cdots i_N} a_{j_1 i_1} a_{j_2 i_2}^{j_1}\cdots a_{j_{N-1} i_{N-1}}^{j_1\cdots j_{N-2}} \right) \nonumber \\
   & & \left( \sum_{l_1\cdots l_{N-1}} \mathcal{C}_{l_1\cdots l_{N-1} i_N} a_{j_1 l_1} a_{j_2 l_2}^{j_1}\cdots a_{j_{N-1} l_{N-1}}^{j_1\cdots j_{N-2}} \right), \nonumber \\
\end{eqnarray}
the equivalence of the definitions of $\rho_{12\cdots m}$ given in Eq.(\ref{e7}) and Eq.(\ref{e18}) and the elements of
$A_1,A_{m-1}^{(j_1\cdots j_{m-2})}, \; m=3,4,\cdots N,$ given by Eq.s(\ref{e60},\;\ref{e61}), and applying the procedure as follows.\\

First, let $3\leq m \leq N$, we take the sum of the index $i_m$ in the $m$ subsystem to get two terms, one with $i_m=0$ and the other with $i_m=1,2,3.$
 After that, we take the sum of the indexes $(i, l)_{m-1}$ of the $m-1$ subsystem to get eight terms. Using Eq.(\ref{e56}) and $a_{j_{m-1} 0}^{(j_1\cdots j_{m-2})}=\frac{1}{\sqrt{2}},$ the four terms containing
$\frac{1}{\sqrt{2}}\sum_{j_{m-1}}^{2} a_{j_{m-1} i_{m-1}}^{(j_1\cdots j_{m-2})}, \; (i,l)_{m-1}=1,2,3$ go to zero
when the sum of $j_{m-1}$ is taken. The  {other} remaining terms,  {two contain} $\hat{e}^{(j_1\cdot j_{m-2})}$ vectors
and the  {other} two terms  {are} without $\hat{e}^{(j_1\cdot j_{m-2})}$ vectors. The terms with $\hat{e}^{(j_1\cdot j_{m-2})}$
vectors constitute $G^{(j_1\cdots j_{m-2})}$ conditional matrix, which is the sum of product of
$\vec{s}^{(m-1)},\; T^{(m-1m)},\;\vec{y}^{(m-1)},\; X^{(m-1m)},\; \vec{y}^{(m-1,j_k)},\; X^{(m-1m,j_k)},\;\vec{y}^{(m-1,\{j_{k_1} j_{k_2}\cdots j_{k_M}\})},$\\
$X^{(m-1m,j_k,\{j_{k_1} j_{k_2}\cdots j_{k_M}\})}, \;\vec{z}^{(m-1,j_k)}, \; W^{(m-1m,j_k)}, \;\vec{z}^{(m-1,\{j_{k_1}, j_{k_2},\cdots,j_{k_M}\})}, \; W^{(m-1m,
\{j_{k_1}, j_{k_2},\cdots,j_{k_M}\})}$
vectors and matrices, as defined below, with the transpose of them.
After taking the sum of indexes in  {the} $m-2$ subsystem, the terms without $\hat{e}^{(j_1\cdot j_{m-2})}$ vectors
produce again terms containing $\hat{e}^{(j_1\cdot j_{m-3})}$ and
terms without any vectors. The terms contain $\hat{e}^{(j_1\cdot j_{m-3})}$  vectors constitute $G^{(j_1\cdots j_{m-3})}$
conditional matrix, which is the sum of the product of
$\vec{s}^{(m-2)},\; T^{(m-2m-1)},\;\vec{y}^{(m-2)},\; X^{(m-2m-1)},\; \vec{y}^{(m-2,j_k)},\; X^{(m-2m-1,j_k)},\;\vec{y}^{(m-2,\{j_{k_1} j_{k_2}\cdots j_{k_M}\})},$\\
$X^{(m-2m-1,\{j_{k_1} j_{k_2}\cdots j_{k_M}\})},
\;\vec{z}^{(m-2,j_k)},\; W^{(m-2m-1,j_k)}, \;\vec{z}^{(m-2,\{j_{k_1}, j_{k_2},\cdots,j_{k_M}\})},$\\
$ W^{(m-2m-1,\{j_{k_1}, j_{k_2},\cdots,j_{k_M}\})}$
 vectors and matrices with a transpose of them.
We continue taking the sum of indexes in every subsystem, in decreasing order, to get a different term with and without
vectors. Every term with vectors constitutes  {a} conditional matrix of a subsystem until to reach the terms
contain unconditional matrix $G=\vec{s}^{(1)}(\vec{s}^{(1)})^t +T^{(1m)}(T^{(1m)})^t$ and others contain only $1+||\vec{s}^{(m)}||^2.$
The  $G^{(j_1j_2\cdots j_{m-2})},\; 3 \leq m \leq N$ can be constructed as follows, first defining\\

$\vec{y}^{(m-1)}_{i_{m-1}}=\sum_{i_1=1}^{3} \hat{e}_{i_1} t_{i_1i_{m-1}},\; X_{i_{m-1}i_m}^{(m-1m)}=\sum_{i_1}^{3} \hat{e}_{i_1} t_{i_1i_{m-1} i_m},\; m=3,\cdots, N,$\\

\noindent for $1\leq k \leq m-3,$\\

$\vec{y}^{(m-1,j_k)}_{i_{m-1}}=\sum_{i_{k+1}}^{3} \hat{e}_{i_{k+1}}^{(j_1\cdots j_k)} t_{i_{k+1}i_{m-1}},$\\

$X_{i_{m-1}i_m}^{(m-1m,j_k)}=\sum_{i_{k+1}}^{3} \hat{e}_{i_{k+1}}^{(j_1\cdots j_k)} t_{i_{k+1}i_{m-1}i_{m}},$\\

$\vec{z}^{(m-1,j_k)}_{i_{m-1}}=\sum_{i_1i_{k+1}}^{3} \hat{e}_{i_1}\hat{e}_{i_{k+1}}^{(j_1\cdots j_k)} t_{i_1i_{k+1}i_{m-1}},$\\

$W_{i_{m-1}i_m}^{(m-1m,j_k)}=\sum_{i_1i_{k+1}}^{3} \hat{e}_{i_1}\hat{e}_{i_{k+1}}^{(j_1\cdots j_k)} t_{i_1i_{k+1}i_{m-1}i_{m}}.$\\

\noindent For every subset $\{k_1,k_2,\cdots,k_M\}$ of $\{1,2,3,\cdots, m-3\},$ and $2\leq M \leq m-3.$\\

$\vec{y}^{(m-1,\{k_1,k_2,\cdots,k_M\})}_{i_{m-1}}=\sum_{i_{k_1+1}i_{k_2+1}\cdots i_{k_M+1}}^{3}  \hat{e}_{i_{k_1+1}}^{(j_1\cdots j_{k_1})} \hat{e}_{i_{k_2+1}}^{(j_1\cdots j_{k_2})}\cdots \hat{e}_{i_{k_M+1}}^{(j_1\cdots j_{k_M})}t_{i_{k_1+1}i_{k_2+1}\cdots i_{k_M+1}i_{m-1}},$\\

$X_{i_{m-1}i_m}^{(m-1m,\{k_1,k_2,\cdots,k_M\})}=\sum_{i_{k_1+1}i_{k_2+1}\cdots i_{k_M+1}}^{3}  \hat{e}_{i_{k_1+1}}^{(j_1\cdots j_{k_1})} \hat{e}_{i_{k_2+1}}^{(j_1\cdots j_{k_2})}\cdots \hat{e}_{i_{k_M+1}}^{(j_1\cdots j_{k_M})}t_{i_{k_1+1}i_{k_2+1}\cdots i_{k_M+1}i_{m-1}i_m},$\\

$\vec{z}^{(m-1,\{k_1,k_2,\cdots,k_M\})}_{i_{m-1}}=\sum_{i_1i_{k_1+1}i_{k_2+1}\cdots i_{k_M+1}}^{3}  \hat{e}_{i_1} \hat{e}_{i_{k_1+1}}^{(j_1\cdots j_{k_1})} \hat{e}_{i_{k_2+1}}^{(j_1\cdots j_{k_2})}\cdots \hat{e}_{i_{k_M+1}}^{(j_1\cdots j_{k_M})}t_{i_1i_{k_1+1}i_{k_2+1}\cdots i_{k_M+1}i_{m-1}},$\\

$W_{i_{m-1}i_m}^{(m-1m,\{k_1,k_2,\cdots,k_M\})}=\sum_{i_1i_{k_1+1}i_{k_2+1}\cdots i_{k_M+1}}^{3}  \hat{e}_{i_1} \hat{e}_{i_{k_1+1}}^{(j_1\cdots j_{k_1})} \hat{e}_{i_{k_2+1}}^{(j_1\cdots j_{k_2})}\cdots \hat{e}_{i_{k_M+1}}^{(j_1\cdots j_{k_M})}t_{i_1i_{k_1+1}i_{k_2+1}\cdots i_{k_M+1}i_{m-1}i_m}.$\\

\noindent Where $\vec{s}^{(m-1)}$ and $T^{(m-1m)}$ take  $+$ sign,\; $\vec{y}^{(m-1)}$ and $X^{(m-1m)}$ take $(-1)^{j_1-1}.$  \\

 $\vec{y}^{(m-1,j_k)}$ and $X^{(m-1m,j_k)}$ take $(-1)^{j_{k+1}-1}; \;$ $\vec{z}^{(m-1,j_k)}$ and $W^{(m-1m,j_k)}$ take $(-1)^{j_1-1}(-1)^{j_{k+1}-1}$ for $k=1,2, \cdots m-3.$\\

 $\vec{y}^{(m-1,\{k_1,k_2,\cdots,k_M\})}$ and $X^{(m-1m,\{k_1,k_2,\cdots,k_M\})}$ take $(-1)^{j_{k_1+1}-1} (-1)^{j_{k_2+1}-1}\cdots(-1)^{j_{k_M+1}-1},$\\

 $\vec{z}^{(m-1,\{k_1,k_2,\cdots,k_M\})}$ and $W^{(m-1m,\{k_1,k_2,\cdots,k_M\})}$  take  $(-1)^{j_1-1}(-1)^{j_{k_1+1}-1} (-1)^{j_{k_2+1}-1}\cdots(-1)^{j_{k_M+1}-1};$ for every subset $\{k_1,k_2,\cdots,k_M\}$ of $\{1,2,3,\cdots, m-3\},\;$  $2\leq M \leq m-3,$ and $j_{k_i},\;j_k=1,2;$ for $j_{k_i},\;k=1,2,3,\cdots, m-3.$\\

So, we get every term with a corresponding sign as follows.\\

\noindent Terms with $+$ sign,\\

$\vec{s}^{(m-1)}(\vec{s}^{(m-1)})^t+ T^{(m-1m)}(T^{(m-1m)})^t,$\\

$\vec{y}^{(m-1)}(\vec{y}^{(m-1)})^t+ X^{(m-1m)}(X^{(m-1m)})^t,$\\

$\vec{y}^{(m-1,j_k)}(\vec{y}^{(m-1,j_k)})^t+ X^{(m-1m,j_k)}(X^{(m-1m,j_k)})^t,$\\

$\vec{z}^{(m-1,j_k)}(\vec{z}^{(m-1,j_k)})^t+ W^{(m-1m,j_k)}(W^{(m-1m,j_k)})^t,$\\

$\vec{y}^{(m-1,\{j_{k_1},j_{k_2},\cdots,j_{k_M}\})}(\vec{y}^{(m-1,\{j_{k_1},j_{k_2},\cdots,j_{k_M}\})})^t+X^{(m-1m,\{j_{k_1},j_{k_2},\cdots,j_{k_M}\})} (X^{(m-1m,\{j_{k_1},j_{k_2},\cdots,j_{k_M}\})})^t,$\\

$\vec{z}^{(m-1,\{j_{k_1},j_{k_2},\cdots,j_{k_M}\})} (\vec{z}^{(m-1,\{j_{k_1},j_{k_2},\cdots,j_{k_M}\})})^t+W^{(m-1m,\{j_{k_1},j_{k_2},\cdots,j_{k_M}\})} (W^{(m-1m,\{j_{k_1},j_{k_2},\cdots,j_{k_M}\})})^t,$\\

\noindent Terms multiply by  $(-1)^{j_1-1},\; j_1=1,2,$ \\

$\vec{s}^{(m-1)}(\vec{y}^{(m-1)})^t+ T^{(m-1m)}(X^{(m-1m)})^t+\vec{y}^{(m-1)}(\vec{s}^{(m-1)})^t+ X^{(m-1m)}(T^{(m-1m)})^t,$\\

\noindent Terms multiply by  $(-1)^{j_{k+1}-1},\; j_{k+1}=1,2,$ \\

$\vec{s}^{(m-1)}(\vec{y}^{(m-1,j_k)})^t+ T^{(m-1m)}(X^{(m-1m,j_k)})^t+\vec{y}^{(m-1,j_k)}(\vec{s}^{(m-1)})^t+ X^{(m-1m,j_k)}(T^{(m-1m)})^t,$\\

$\vec{y}^{(m-1)}(\vec{z}^{(m-1,j_k)})^t+ X^{(m-1m)}(W^{(m-1m,j_k)})^t+\vec{z}^{(m-1,j_k)}(\vec{y}^{(m-1)})^t+ W^{(m-1m,j_k)}(X^{(m-1m)})^t,$\\

\noindent Terms multiply by  $(-1)^{j_1-1}(-1)^{j_{k+1}-1},$ \\

$\vec{s}^{(m-1)}(\vec{z}^{(m-1,j_k)})^t+ T^{(m-1m)}(W^{(m-1m,j_k)})^t+\vec{z}^{(m-1,j_k)}(\vec{s}^{(m-1)})^t+ W^{(m-1m,j_k)}(T^{(m-1m)})^t,$\\

$\vec{y}^{(m-1)}(\vec{y}^{(m-1,j_k)})^t+ X^{(m-1m)}(X^{(m-1m,j_k)})^t+\vec{y}^{(m-1,j_k)}(\vec{y}^{(m-1)})^t+ X^{(m-1m,j_k)}(X^{(m-1m)})^t,$\\

\noindent Terms multiply by $(-1)^{j_{k_1+1}-1} (-1)^{j_{k_2+1}-1}\cdots (-1)^{j_{k_M+1}-1},$ \\

$\vec{s}^{(m-1)}(\vec{y}^{(m-1,\{j_{k_1},j_{k_2},\cdots,j_{k_M}\})})^t+ T^{(m-1m)}(X^{(m-1m,\{j_{k_1},j_{k_2},\cdots,j_{k_M}\})})^t+\vec{y}^{(m-1,\{j_{k_1},j_{k_2},\cdots,j_{k_M}\})}(\vec{s}^{(m-1)})^t+ X^{(m-1m,\{j_{k_1},j_{k_2},\cdots,j_{k_M}\})}(T^{(m-1m)})^t,$\\

$\vec{y}^{(m-1)}(\vec{z}^{(m-1,\{j_{k_1},j_{k_2},\cdots,j_{k_M}\})})^t+ X^{(m-1m)}(W^{(m-1m,\{j_{k_1},j_{k_2},\cdots,j_{k_M}\})})^t+\vec{z}^{(m-1,\{j_{k_1},j_{k_2},\cdots,j_{k_M}\})}(\vec{y}^{(m-1)})^t+ W^{(m-1m,\{j_{k_1},j_{k_2},\cdots,j_{k_M}\})}(X^{(m-1m)})^t,$\\

\noindent Terms multiply by $(-1)^{j_1-1}(-1)^{j_{k_1+1}-1} (-1)^{j_{k_2+1}-1}\cdots (-1)^{j_{k_M+1}-1},$ \\

$\vec{s}^{(m-1)}(\vec{z}^{(m-1,\{j_{k_1},j_{k_2},\cdots,j_{k_M}\})})^t+ T^{(m-1m)}(W^{(m-1m,\{j_{k_1},j_{k_2},\cdots,j_{k_M}\})})^t+\vec{z}^{(m-1,\{j_{k_1},j_{k_2},\cdots,j_{k_M}\})}(\vec{s}^{(m-1)})^t+ W^{(m-1m,\{j_{k_1},j_{k_2},\cdots,j_{k_M}\})}(T^{(m-1m)})^t,$\\

$\vec{y}^{(m-1)}(\vec{y}^{(m-1,\{j_{k_1},j_{k_2},\cdots,j_{k_M}\})})^t+ X^{(m-1m)}(X^{(m-1m,\{j_{k_1},j_{k_2},\cdots,j_{k_M}\})})^t+\vec{y}^{(m-1,\{j_{k_1},j_{k_2},\cdots,j_{k_M}\})}(\vec{y}^{(m-1)})^t+ X^{(m-1m,\{j_{k_1},j_{k_2},\cdots,j_{k_M}\})}(X^{(m-1m)})^t,$\\

\noindent Terms multiply by $(-1)^{j_{k+1}-1}(-1)^{j_{k_1+1}-1} (-1)^{j_{k_2+1}-1}\cdots (-1)^{j_{k_M+1}-1},$\\

$\vec{y}^{(m-1,j_k)}(\vec{y}^{(m-1,\{j_{k_1},j_{k_2},\cdots,j_{k_M}\})})^t+ X^{(m-1m,j_k)}(X^{(m-1m,\{j_{k_1},j_{k_2},\cdots,j_{k_M}\})})^t+\vec{y}^{(m-1,\{j_{k_1},j_{k_2},\cdots,j_{k_M}\})}(\vec{y}^{(m-1,j_k)})^t+ X^{(m-1m,\{j_{k_1},j_{k_2},\cdots,j_{k_M}\})}(X^{(m-1m,j_k)})^t,$\\

\noindent Terms multiply by $(-1)^{j_1-1}(-1)^{j_{k+1}-1}(-1)^{j_{k_1+1}-1} (-1)^{j_{k_2+1}-1}\cdots (-1)^{j_{k_M+1}-1},$\\

$\vec{y}^{(m-1,j_k)}(\vec{z}^{(m-1,\{j_{k_1},j_{k_2},\cdots,j_{k_M}\})})^t+ X^{(m-1m,j_k)}(W^{(m-1m,\{j_{k_1},j_{k_2},\cdots,j_{k_M}\})})^t+\vec{z}^{(m-1,\{j_{k_1},j_{k_2},\cdots,j_{k_M}\})}(\vec{y}^{(m-1,j_k)})^t+ W^{(m-1m,\{j_{k_1},j_{k_2},\cdots,j_{k_M}\})}(X^{(m-1m,j_k)})^t,$\\

\noindent for the subsets $\{k_1,k_2,\cdots,k_M\} \neq \{q_1,q_2,\cdots,q_L\}$ of $\{1,2,3,\cdots,m-3\}$ and $2\leq M \leq m-3,\; 2\leq L \leq m-3,$\\

\noindent Terms multiply by $(-1)^{j_{q_1+1}-1} (-1)^{j_{q_2+1}-1}\cdots (-1)^{j_{q_L+1}-1}(-1)^{j_{k_1+1}-1} (-1)^{j_{k_2+1}-1}\cdots (-1)^{j_{k_M+1}-1},$\\

$\vec{y}^{(m-1,\{j_{q_1},j_{q_2},\cdots,j_{q_L}\})}(\vec{y}^{(m-1,\{j_{k_1},j_{k_2},\cdots,j_{k_M}\})})^t+ X^{(m-1m,\{j_{q_1},j_{q_2},\cdots,j_{q_L}\})}(X^{(m-1m,\{j_{k_1},j_{k_2},\cdots,j_{k_M}\})})^t+\vec{y}^{(m-1,\{j_{k_1},j_{k_2},\cdots,j_{k_M}\})}(\vec{y}^{(m-1,\{j_{q_1},j_{q_2},\cdots,j_{q_L}\})})^t+ X^{(m-1m,\{j_{k_1},j_{k_2},\cdots,j_{k_M}\})}(X^{(m-1m,\{j_{q_1},j_{q_2},\cdots,j_{q_L}\})})^t,$\\

$\vec{z}^{(m-1,\{j_{q_1},j_{q_2},\cdots,j_{q_L}\})}(\vec{z}^{(m-1,\{j_{k_1},j_{k_2},\cdots,j_{k_M}\})})^t+ W^{(m-1m,\{j_{q_1},j_{q_2},\cdots,j_{q_L}\})}(W^{(m-1m,\{j_{k_1},j_{k_2},\cdots,j_{k_M}\})})^t+\vec{z}^{(m-1,\{j_{k_1},j_{k_2},\cdots,j_{k_M}\})}(\vec{z}^{(m-1,\{j_{q_1},j_{q_2},\cdots,j_{q_L}\})})^t+ W^{(m-1m,\{j_{k_1},j_{k_2},\cdots,j_{k_M}\})}(W^{(m-1m,\{j_{q_1},j_{q_2},\cdots,j_{q_L}\})})^t,$\\
\noindent Terms multiply by $(-1)^{j_1-1}(-1)^{j_{q_1+1}-1} (-1)^{j_{q_2+1}-1}\cdots (-1)^{j_{q_L+1}-1}(-1)^{j_{k_1+1}-1} (-1)^{j_{k_2+1}-1}\cdots (-1)^{j_{k_M+1}-1},$\\

$\vec{y}^{(m-1,\{j_{q_1},j_{q_2},\cdots,j_{q_L}\})}(\vec{z}^{(m-1,\{j_{k_1},j_{k_2},\cdots,j_{k_M}\})})^t+ X^{(m-1m,\{j_{q_1},j_{q_2},\cdots,j_{q_L}\})}(W^{(m-1m,\{j_{k_1},j_{k_2},\cdots,j_{k_M}\})})^t+\vec{z}^{(m-1,\{j_{k_1},j_{k_2},\cdots,j_{k_M}\})}(\vec{y}^{(m-1,\{j_{q_1},j_{q_2},\cdots,j_{q_L}\})})^t+ W^{(m-1m,\{j_{k_1},j_{k_2},\cdots,j_{k_M}\})}(X^{(m-1m,\{j_{q_1},j_{q_2},\cdots,j_{q_L}\})})^t,$\\

Now, the $G,$ and $G^{(j_1j_2\cdots j_{m-2})},\; 3 \leq m \leq N$ and $j_1, j_2, \cdots, j_{m-2}=1,2$ are,
\begin{eqnarray} \label{e65}
G  &=& \vec{s}^{(1)}(\vec{s}^{(1)})^t+ T^{(1m)}(T^{(1m)})^t,  \nonumber \\
G^{(j_1j_2\cdots j_{m-2})} &=& \frac{1}{2^{m-2}}\left\{ \vec{s}^{(m-1)}(\vec{s}^{(m-1)})^t+ T^{(m-1m)}(T^{(m-1m)})^t+\vec{y}^{(m-1)}(\vec{y}^{(m-1)})^t+X^{(m-1m)}(X^{(m-1m)})^t\right. \nonumber \\
& & +\left.+(-1)^{j_1-1}\left[\vec{s}^{(m)}(\vec{y}^{(m)})^t+ T^{(mN)}(X^{(mN)})^t+\vec{y}^{(m)}(\vec{s}^{(m)})^t+ X^{(mN)}(T^{(mN)})^t  \right]\right.  \nonumber \\
& & +\left. \sum_{k=1}^{m-3}\left[\vec{y}^{(m-1,j_k)}(\vec{y}^{(m-1,j_k)})^t +X^{(m-1m,j_k)}(X^{(m-1m,j_k)})^t \right. \right.  \nonumber \\
& &\left. \left. +\vec{z}^{(m-1,j_k)}(\vec{z}^{(m-1,j_k)})^t +W^{(m-1m,j_k)}(W^{(m-1m,j_k)})^t\right.\right.\nonumber \\
& &\left. \left. +(-1)^{j_{k+1}-1} \left[\vec{s}^{(m-1)}(\vec{y}^{(m-1,j_k)})^t+ T^{(m-1m)}(X^{(m-1m,j_k)})^t+\vec{y}^{(m-1,j_k)}(\vec{s}^{(m-1)})^t
\right. \right. \right. \nonumber\\
& & \left. \left. \left.+ X^{(m-1m,j_k)}(T^{(m-1m)})^t+\vec{y}^{(m-1)}(\vec{z}^{(m-1,j_k)})^t+ X^{(m-1m)}(W^{(m-1m,j_k)})^t
\right. \right.\right. \nonumber\\
& &  \left.\left.\left.  +\vec{z}^{(m-1,j_k)}(\vec{y}^{(m-1)})^t+ W^{(m-1m,j_k)}(X^{(m-1m)})^t+(-1)^{j_1-1}\left[\vec{s}^{(m-1)}(\vec{z}^{(m-1,j_k)})^t \right.\right. \right.\right. \nonumber\\
& & \left. \left. \left.\left. + T^{(m-1m)}(W^{(m-1m,j_k)})^t+\vec{z}^{(m-1,j_k)}(\vec{s}^{(m-1)})^t+ W^{(m-1m,j_k)}(T^{(m-1m)})^t\right. \right.\right.\right. \nonumber\\
& & \left. \left. \left.\left.+\vec{y}^{(m-1)}(\vec{y}^{(m-1,j_k)})^t+ X^{(m-1m)}(X^{(m-1m,j_k)})^t+\vec{y}^{(m-1,j_k)}(\vec{y}^{(m-1)})^t\right. \right.\right.\right. \nonumber\\
& & \left.\left.\left. \left. + X^{(m-1m,j_k)}(X^{(m-1m)})^t \right] \right]\right]\right. \nonumber\\
& &\left. +\sum_{M=2}^{m-3} \sum_{\{k_1,k_2,\cdots,k_M\}}  \left[\vec{y}^{(m-1,\{k_1,k_2,\cdots,k_M\})}(\vec{y}^{(m-1,\{k_1,k_2,\cdots,k_M\})})^t\right.\right.\nonumber \\
& & \left. \left. + X^{(m-1m,\{k_1,k_2,\cdots,k_M\})}(X^{(m-1m,\{k_1,k_2,\cdots,k_M\})})^t
+\vec{z}^{(m-1,\{k_1,k_2,\cdots,k_M\})}(\vec{z}^{(m-1,\{k_1,k_2,\cdots,k_M\})})^t \right. \right. \nonumber \\
& &+\left. \left. W^{(m-1m,\{k_1,k_2,\cdots,k_M\})}(W^{(m-1m,\{k_1,k_2,\cdots,k_M\})})^t \right.\right. \nonumber\\
& &\left. \left. +(-1)^{j_{k_1+1}-1} (-1)^{j_{k_2+1}-1}\cdots (-1)^{j_{k_M+1}-1}\left[\vec{s}^{(m-1)}(\vec{y}^{(m-1,\{j_{k_1},j_{k_2},\cdots,j_{k_M}\})})^t\right. \right.\right. \nonumber \\
& & \left. \left. \left.+ T^{(m-1m)}(X^{(m-1m,\{j_{k_1},j_{k_2},\cdots,j_{k_M}\})})^t
+\vec{y}^{(m-1,\{j_{k_1},j_{k_2},\cdots,j_{k_M}\})}(\vec{s}^{(m-1)})^t \right. \right.\right. \nonumber\\
& & \left. \left. \left.+ X^{(m-1m,\{j_{k_1},j_{k_2},\cdots,j_{k_M}\})}(T^{(m-1m)})^t
+\vec{y}^{(m-1)}(\vec{z}^{(m-1,\{j_{k_1},j_{k_2},\cdots,j_{k_M}\})})^t \right. \right.\right. \nonumber\\
& &\left. \left.\left. + X^{(m-1m)}(W^{(m-1m,\{j_{k_1},j_{k_2},\cdots,j_{k_M}\})})^t
+\vec{z}^{(m-1,\{j_{k_1},j_{k_2},\cdots,j_{k_M}\})}(\vec{y}^{(m-1)})^t\right. \right.\right. \nonumber\\
& &\left. \left.\left. + W^{(m-1m,\{j_{k_1},j_{k_2},\cdots,j_{k_M}\})}(X^{(m-1m)})^t +(-1)^{j_1-1}\left[\vec{s}^{(m-1)}(\vec{z}^{(m-1,\{j_{k_1},j_{k_2},\cdots,j_{k_M}\})})^t \right. \right.\right.\right. \nonumber\\
& &\left. \left.\left. \left.+ T^{(m-1m)}(W^{(m-1m,\{j_{k_1},j_{k_2},\cdots,j_{k_M}\})})^t
+\vec{z}^{(m-1,\{j_{k_1},j_{k_2},\cdots,j_{k_M}\})}(\vec{s}^{(m-1)})^t \right.\right.\right.\right. \nonumber\\
& &\left. \left.\left. \left. + W^{(m-1m,\{j_{k_1},j_{k_2},\cdots,j_{k_M}\})}(T^{(m-1m)})^t
+\vec{y}^{(m-1)}(\vec{y}^{(m-1,\{j_{k_1},j_{k_2},\cdots,j_{k_M}\})})^t \right.\right.\right.\right. \nonumber\\
& &\left. \left.\left.\left.+ X^{(m-1m)}(X^{(m-1m,\{j_{k_1},j_{k_2},\cdots,j_{k_M}\})})^t
+\vec{y}^{(m-1,\{j_{k_1},j_{k_2},\cdots,j_{k_M}\})}(\vec{y}^{(m-1)})^t\right.\right.\right.\right. \nonumber\\
& &\left. \left.\left.\left. + X^{(m-1m,\{j_{k_1},j_{k_2},\cdots,j_{k_M}\})}(X^{(m-1m)})^t \right]\right]\right]+F \right\}.\nonumber\\
\end{eqnarray}
The matrix $F$ is defined as,
\begin{eqnarray}\label{e66}
F & = &\sum_{k=1}^{m-3}\sum_{M=2}^{m-3} \sum_{\{k_1,\cdots,k_M\}} (-1)^{j_{k+1}-1}(-1)^{j_{k_1+1}-1} (-1)^{j_{k_2+1}-1}\cdots(-1)^{j_{k_M+1}-1}\nonumber\\
& & \times \left[\vec{y}^{(m-1,j_k)}(\vec{y}^{(m-1,\{j_{k_1},j_{k_2},\cdots,j_{k_M}\})})^t+ X^{(m-1m,j_k)}(X^{(m-1m,\{j_{k_1},j_{k_2},\cdots,j_{k_M}\})})^t\right.  \nonumber\\
& &\left.+\vec{y}^{(m-1,\{j_{k_1},j_{k_2},\cdots,j_{k_M}\})}(\vec{y}^{(m-1,j_k)})^t + X^{(m-1m,\{j_{k_1},j_{k_2},\cdots,j_{k_M}\})}(X^{(m-1m,j_k)})^t \right.  \nonumber\\
& & \left.+(-1)^{j_1-1} \left[\vec{y}^{(m-1,j_k)}(\vec{z}^{(m-1,\{j_{k_1},j_{k_2},\cdots,j_{k_M}\})})^t + X^{(m-1m,j_k)}(W^{(m-1m,\{j_{k_1},j_{k_2},\cdots,j_{k_M}\})})^t \right.\right.  \nonumber\\
& & \left.\left.+\vec{z}^{(m-1,\{j_{k_1},j_{k_2},\cdots,j_{k_M}\})}(\vec{y}^{(m-1,j_k)})^t+ W^{(m-1m,\{j_{k_1},j_{k_2},\cdots,j_{k_M}\})}(X^{(m-1m,j_k)})^t \right]\right]  \nonumber\\
& & +\sum_{M=2}^{m-3} \sum_{L=2}^{m-3}\sum_{\{k_1,k_2,\cdots,k_M\}}\sum_{\{q_1,q_2,\cdots,q_L\}\neq \{k_1,k_2,\cdots,k_M\}} (-1)^{j_{q_1+1}-1}\cdots (-1)^{j_{q_L+1}-1}(-1)^{j_{k_1+1}-1} \cdots \nonumber\\
& &\times (-1)^{j_{k_M+1}-1} \left[\vec{y}^{(m-1,\{j_{q_1},j_{q_2},\cdots,j_{q_L}\})}(\vec{y}^{(m-1,\{j_{k_1},j_{k_2},\cdots,j_{k_M}\})})^t \right.\nonumber\\
& & \left.+X^{(m-1m,\{j_{q_1},j_{q_2},\cdots,j_{q_L}\})}(X^{(m-1m,\{j_{k_1},j_{k_2},\cdots,j_{k_M}\})})^t \right.\nonumber\\
& & \left. +\vec{y}^{(m-1,\{j_{k_1},j_{k_2},\cdots,j_{k_M}\})}(\vec{y}^{(m-1,\{j_{q_1},j_{q_2},\cdots,j_{q_L}\})})^t\right. \nonumber\\
& & \left.+X^{(m-1m,\{j_{k_1},j_{k_2},\cdots,j_{k_M}\})}(X^{(m-1m,\{j_{q_1},j_{q_2},\cdots,j_{q_L}\})})^t \right. \nonumber\\
& &\left. + \vec{z}^{(m-1,\{j_{q_1},j_{q_2},\cdots,j_{q_L}\})}(\vec{z}^{(m-1,\{j_{k_1},j_{k_2},\cdots,j_{k_M}\})})^t\right. \nonumber\\
& &\left.+W^{(m-1m,\{j_{q_1},j_{q_2},\cdots,j_{q_L}\})}(W^{(m-1m,\{j_{k_1},j_{k_2},\cdots,j_{k_M}\})})^t\right. \nonumber\\
& &\left.+\vec{z}^{(m-1,\{j_{k_1},j_{k_2},\cdots,j_{k_M}\})}(\vec{z}^{(m-1,\{j_{q_1},j_{q_2},\cdots,j_{q_L}\})})^t\right.\nonumber\\
& &\left.+W^{(m-1m,\{j_{k_1},j_{k_2},\cdots,j_{k_M}\})}(W^{(m-1m,\{j_{q_1},j_{q_2},\cdots,j_{q_L}\})})^t\right.\nonumber\\
& & \left.+ (-1)^{j_1-1}\left[\vec{y}^{(m-1,\{j_{q_1},j_{q_2},\cdots,j_{q_L}\})}(\vec{z}^{(m-1,\{j_{k_1},j_{k_2},\cdots,j_{k_M}\})})^t \right.\right.\nonumber\\
& & \left.\left.+ X^{(m-1m,\{j_{q_1},j_{q_2},\cdots,j_{q_L}\})}(W^{(m-1m,\{j_{k_1},j_{k_2},\cdots,j_{k_M}\})})^t \right.\right.\nonumber\\
& & \left.\left.+\vec{z}^{(m-1,\{j_{k_1},j_{k_2},\cdots,j_{k_M}\})}(\vec{y}^{(m-1,\{j_{q_1},j_{q_2},\cdots,j_{q_L}\})})^t \right.\right.\nonumber\\
& & \left.\left.+ W^{(m-1m,\{j_{k_1},j_{k_2},\cdots,j_{k_M}\})}(X^{(m-1m,\{j_{q_1},j_{q_2},\cdots,j_{q_L}\})})^t\right]\right]\nonumber\\
\end{eqnarray}
Now, the Eq. (\ref{e64}) becomes
\begin{eqnarray}\label{e67}
||\mathcal{C}\times_1 A_1 \times_2 \cdots \times_{N-1} A_{N-1}^{(j_1j_2\cdots j_{N-2})}||^2  &=& \frac{1}{2^N}\left\{1+||\vec{s}^{(N)}||^2 + \hat{e} G \hat{e}^t  \right. \nonumber \\
& & +\left.  \sum_{m=3}^{N}\sum_{j_1\cdots j_m=1}^{2}\hat{e}^{(j_1\cdots j_{m-2})}\;G^{(j_1\cdots j_{m-2})}\;(\hat{e}^{(j_1\cdots j_{m-2})})^t  \right\}. \nonumber \\
\end{eqnarray}
In Eq.(\ref{e67}) the last two terms {depend} on matrices $A_1,$ and $A_{m-1}^{(j_1\cdots j_{m-2})}$ while others are determined by the state $\rho_{1\cdots N}.$ Therefore, to maximize
$||\mathcal{C}\times_1 A_1 \times_2 A_2^{(j_1)}\times_3 \cdots A_{N-1}^{(j_1j_2\cdots j_{N-2})}||^2$
we take $\hat{e},\; \hat{e}^{(j_1\cdots j_{m-2})}$ to be the eigenvectors of $G, \; G^{(j_1\cdots j_{m-2})}$ corresponding
to its largest eigenvalues $\eta_{max},\;\eta_{max}^{(j_1\cdots j_{m-2})}$ respectively,  so that

\begin{eqnarray}\label{e68}
\max_{A_1,A_2^{(j_1)},\cdots,A_{N-1}^{(j_1j_2\cdots j_{N-2})}}||\mathcal{C}\times_1 A_1 \times_2 A_2^{(j_1)}\times_3\cdots A_{N-1}^{(j_1j_2\cdots j_{N-2})}||^2 &=& \frac{1}{2^N}\left\{1+||\vec{s}^{(N)}||^2+\eta_{max} \right. \nonumber \\
& &  +\left. \sum_{m=3}^{N}\sum_{j_1\cdots j_m=1}^{2}\eta_{max}^{(j_1\cdots j_{m-2})} \right\}. \nonumber \\
\end{eqnarray}
Finally, Eq.(\ref{e68}), Eq.(\ref{e62}) and  Eq.(\ref{e10}) together imply

\begin{eqnarray*}
GD_{A_1;\cdots;A_N}(\rho_{1\cdots N}) & = & \frac{1}{2^N} \left[ \sum_{k=1}^{N-1} ||\vec{s}^{(k)}||^2+\sum_{\{k_1,k_2\}} ||T^{\{k_1,k_2\}}||^2+\cdots+\sum_{\{k_1,\cdots,k_M\}}  ||T^{\{k_1,k_2,\cdots,k_M\}}||^2 \right. \nonumber \\
& &+\left. \cdots+ ||\mathcal{T}||^2-\eta_{max}-\sum_{m=3}^{N} \sum_{j_1\cdots j_m=1}^{2}\eta_{max}^{(j_1\cdots j_{m-2})}
\right].
 \end{eqnarray*}
where $\eta_{max},\;\eta_{max}^{(j_1\cdots j_{m-2})},\; j_k=1,2,\; k=1,\cdots, m-2$ are the largest eigenvalues of unconditional matrix $G$ and conditional matrices  $G^{(j_1\cdots j_{m-2})}$ respectively, thus completing the proof.\\

If we consider the following family of $N$-qubit states \cite{zhu},

\begin{equation}\label{e69}
  \rho=\frac{1}{2^N} (I+ \sum_{j=1}^{3} c_j \sigma_j \otimes \cdots \otimes \sigma_j).
\end{equation}
According to Eq.(\ref{e18}) all terms between $I$ identity term and $\mathcal{T}$ total correlation tensor (last term) are zero. So, the $GD$ depends on the $\mathcal{T}$ tensor only.\\

First, let $N=3,$ the state is
\begin{equation}\label{e70}
  \rho=\frac{1}{2^3} (I+ \sum_{j=1}^{3} c_j \sigma_j \otimes \sigma_j \otimes \sigma_j).
\end{equation}
Since all terms are zero except two terms $I$ and $\mathcal{T},$ so we have
\begin{eqnarray}\label{e71}
  ||\mathcal{C}||^2 & = & \frac{1}{2^3}\left[1+||\mathcal{T}||^2\right].
\end{eqnarray}
{To} get the norm of $\mathcal{C}\times_1 A_1\times_2 A_2^{(j_1)}$ we use its elementwise definition,
\begin{equation}\label{e72}
  (\mathcal{C}\times_1 A_1\times_2 A_2^{(j_1)})_{j_1j_2i_3}  =
   \sum_{i_1i_2} c_{i_1i_2 i_3} a_{j_1i_1}a_{j_2 i_2}^{(j_1)},  \; j_1=1,2, \; j_2=1,2,
\end{equation}
the equivalence of the definitions of $\rho_{123}$ given in Eq.(\ref{e7}) and Eq.(\ref{e18}) for these class of states and the elements of
$A_1,\; A_2^{(j_1)}$ given by Eq.s(\ref{e29},\ref{e30}). The result is

\begin{eqnarray}\label{e73}
  ||\mathcal{C}\times_1 A_1 \times_2 A_2^{(j_1)}||^2 &=& \frac{1}{2^3}\left[1+ \frac{1}{2^2}\sum_{i_3}^{3}\sum_{i_1 i_2}^{3}\sum_{l_1 l_2}^{3} t_{i_1 i_2 i_3}t_{l_1 l_2 i_3} \sum_{j_1 j_2}^{2} e_{j_1i_1}e_{j_1l_1} e^{(j_1)}_{j_2i_2}e^{(j_1)}_{j_2l_2}\right],
\end{eqnarray}
using Eq.(\ref{e27}), we have
\begin{eqnarray}\label{e74}
  ||\mathcal{C}\times_1 A_1 \times_2 A_2^{(j_1)}||^2 &=& \frac{1}{2^3}\left[1+ \frac{1}{2}\sum_{i_3}^{3}\sum_{i_1 i_2}^{3}\sum_{l_1 l_2}^{3} t_{i_1 i_2 i_3}t_{l_1 l_2 i_3} e_{i_1}e_{l_1} (e^{(1)}_{i_2}e^{(1)}_{l_2}+e^{(1)}_{i_2}e^{(1)}_{l_2})\right].
\end{eqnarray}
we can write $t_{i_1i_2i_3}=\sum_i c_i \delta_{ii_1}\delta_{ii_2}\delta_{ii_3}$ and $t_{l_1l_2i_3}=\sum_l c_l \delta_{ll_1}\delta_{ll_2}\delta_{li_3},$ we get

\begin{eqnarray}\label{e75}
  ||\mathcal{C}\times_1 A_1 \times_2 A_2^{(j_1)}||^2 &=& \frac{1}{2^3}\left[1+ \frac{1}{2}\sum_{i}^{3} c^2_i e^2_i((e^{(1)}_i)^2 +(e^{(2)}_i)^2)\right].
\end{eqnarray}
From Eqs.(\ref{e71}) and (\ref{e75}), we obtain

\begin{eqnarray}\label{e76}
||\mathcal{C}||^2-||\mathcal{C}\times_1 A_1\times_2 A_2^{(j_1)}||^2
&=& \frac{1}{2^3}\left[||\mathcal{T}||^2-\frac{1}{2}\sum_{i}^{3} c^2_i e^2_i((e^{(1)}_i)^2 +(e^{(2)}_i)^2)\right].
\end{eqnarray}
We have $||\hat{e}||^2=e^2_1+e^2_2+e^2_3=1$ and $||\hat{e}^{(j_1)}||^2=(e_1^{(j_1)})^2+(e_2^{(j_1)})^2+(e_3^{(j_1)})^2=1,\; j_1=1,2.$ Let

\begin{equation}\label{e77}
c=\max\{|c_1|,|c_2|,|c_3|\}.
\end{equation}
Then, the second term in RHS of Eq.(\ref{e76}) is

\begin{eqnarray}\label{e78}
& &\frac{1}{2}\left[ c^2_1 e^2_1((e^{(1)}_1)^2 +(e^{(2)}_1)^2)+c^2_2 e^2_2((e^{(1)}_2)^2 +(e^{(2)}_2)^2)+c^2_3 e^2_3((e^{(1)}_3)^2 +(e^{(2)}_3)^2)\right] \nonumber \\
& &\leq \frac{1}{2} |c|^2 \left[ e^2_1((e^{(1)}_1)^2 +(e^{(2)}_1)^2)+e^2_2((e^{(1)}_2)^2 +(e^{(2)}_2)^2)+e^2_3((e^{(1)}_3)^2 +(e^{(2)}_3)^2)\right]=c^2,
\end{eqnarray}
we can get the last equality in Eq. (\ref{e78}) by choosing appropriate vectors $\hat{e}$ and $\hat{e}^{(j_1)}, \; j_1=1,2.$

Therefore, the minimization of Eq.(\ref{e76}) is by taking the maximization of the second term in RHS of it and the $GD_{A_1; A_2; A_3}(\rho_{123})$ is given by

\begin{equation*}
   GD_{A_1;A-2;A_3}(\rho_{123})=\frac{1}{2^3}\left[||\mathcal{T}||^2-c^2\right],
\end{equation*}

and $||\mathcal{T}||^2=c_1^2+c_2^2+c_3^2,$ hence the GD is
\begin{equation}\label{e79}
 GD_{A_1;A_2;A_3}(\rho_{123})=\frac{1}{2^3}\left[ c_1^2+c_2^2+c_3^2-c^2\right].
 \end{equation}
Which is as in ref. \cite{zhu}.

Second, for $N > 3,$ the state is as in Eq.(\ref{e69}).\\
The norm of the tensor $\mathcal{C}$ can be expressed in terms of the norms of the tensors defining $\rho_{12\cdots N}$
by using the equivalence of the definitions of $\rho_{12\cdots N}$ given in Eq.(\ref{e7}) and Eq.(\ref{e18}) for this family of states as

\begin{eqnarray}\label{e80}
  ||\mathcal{C}||^2 & = & \frac{1}{2^N}\left[1+||\mathcal{T}||^2\right]. \nonumber \\
\end{eqnarray}
In order to get the norm of $\mathcal{C}\times_1 A_1\times_2 A_2^{(j_1)}\times \cdots\times_{N-1} A_{N-1}^{(j_1j_2\cdots j_{N-2})}$ we use its elementwise definition,

\begin{eqnarray}\label{e81}
 (\mathcal{C}\times_1 A_1\times_2 A_2^{(j_1)}\times \cdots\times_{N-1} A_{N-1}^{(j_1j_2\cdots j_{N-2})})_{j_1\cdots j_{N-1} i_N}  &=& \sum_{i_1i_2\cdots i_{N-1}} c_{i_1i_2 \cdots i_N} a_{j_1i_1}a_{j_2 i_2}^{(j_1)}\cdots a_{j_{N-1} i_{N-1}}^{(j_1j_2\cdots j_{N-2})}, \nonumber \\
\end{eqnarray}
Let us put $i_k=0,1,2,3$ instead of $i_k=1,2,3,4,\; k=1,\cdots N,$ the norm can be obtained as,

\begin{eqnarray}\label{e82}
  ||\mathcal{C}\times_1 A_1 \times_2 A_2^{(j_1)}\times_3 \cdots \times_{N-1} A_{N-1}^{(j_1j_2\cdots j_{N-2})}||^2 &=& \sum_{j_1\cdots j_{N-1}}^{2} \sum_{i_N=0}^{3}\left(\sum_{i_1\cdots i_{N-1}} \mathcal{C}_{i_1\cdots i_N} a_{j_1 i_1} a_{j_2 i_2}^{j_1}\cdots a_{j_{N-1} i_{N-1}}^{j_1\cdots j_{N-2}} \right) \nonumber \\
   & & \left( \sum_{l_1\cdots l_{N-1}} \mathcal{C}_{l_1\cdots l_{N-1} i_N} a_{j_1 l_1} a_{j_2 l_2}^{j_1}\cdots a_{j_{N-1} l_{N-1}}^{j_1\cdots j_{N-2}} \right), \nonumber \\
\end{eqnarray}
the equivalence of the definitions of $\rho_{12\cdots N}$ given in Eq.(\ref{e7}) and Eq.(\ref{e18}) for these class of states and the elements of $A_1,A_{m-1}^{(j_1\cdots j_{m-2})}, \; m=3,4,\cdots N,$ given by Eq.s(\ref{e60},\;\ref{e61}). The result is

\begin{eqnarray}\label{e83}
  ||\mathcal{C}\times_1 A_1 \times_2 A_2^{(j_1)}\times_3 \cdots \times_{N-1} A_{N-1}^{(j_1j_2\cdots j_{N-2})}||^2 &=&\frac{1}{2^N}\left[1  \right. \nonumber \\
 & & \left. +\frac{1}{2^{N-1}}\sum_{i_N}^{3}\sum_{i_1 i_2 \cdots i_{N-1}}^{3}\sum_{l_1 l_2\cdots l_{N-1} i_N}^{3} t_{i_1 i_2\cdots i_N}t_{l_1 l_2 \cdots l_{N-1} i_N} \right.\nonumber  \\
 & & \left. \times \sum_{j_1\cdots j_{N-1}}^{2} e_{j_1i_1}e_{j_1l_1} e^{(j_1)}_{j_2i_2}e^{(j_1)}_{j_2l_2}\cdots e^{(j_1j_2\cdots j_{N-2})}_{j_{N-1}i_{N-1}}e^{(j_1j_2\cdots j_{N-2})}_{j_{N-1}l_{N-1}} \right],\nonumber \\
\end{eqnarray}
using Eq.(\ref{e58}) for $e_{j_1}$ and $e^{(j_1j_2\cdots j_{N-2})}_{j_{N-1}}$, we have

\begin{eqnarray}\label{e84}
  & &||\mathcal{C}\times_1 A_1 \times_2 A_2^{(j_1)}\times_3 \cdots \times_{N-1} A_{N-1}^{(j_1j_2\cdots j_{N-2})}||^2\nonumber \\
  &=& \frac{1}{2^N}\left[1 + \frac{1}{2^{N-2}}\sum_{i_N}^{3}\sum_{i_1 i_2 \cdots i_{N-1}}^{3} \sum_{l_1 l_2\cdots l_{N-1} i_N}^{3} t_{i_1 i_2\cdots i_N}t_{l_1 l_2 \cdots l_{N-1} i_N} e_{i_1}e_{l_1} \right. \nonumber \\
  & & \times \left. \left(\sum_{j_1}^{2} e^{(j_1)}_{i_2}e^{(j_1)}_{l_2}\left(\cdots \left( \sum_{j_{N-2}}^{2}e^{(j_1j_2\cdots j_{N-2})}_{i_{N-1}} e^{(j_1j_2\cdots j_{N-2})}_{l_{N-1}} \right)\cdots\right)\right) \right],\nonumber  \\
\end{eqnarray}
we can write $t_{i_1i_2 \cdots i_N}=\sum_i c_i \delta_{ii_1}\delta_{ii_2} \cdots \delta_{ii_N}$ and $t_{l_1l_2 \cdots l_{N-1}i_N}=\sum_l c_l \delta_{ll_1}\delta_{ll_2} \cdots \delta_{ll_{N-1}}\delta_{li_N},$ we get

\begin{eqnarray}\label{e85}
  ||\mathcal{C}\times_1 A_1 \times_2 A_2^{(j_1)}||^2 &=& \frac{1}{2^N}\left[1+ \frac{1}{2^{N-2}}\sum_{i}^{3} c^2_i e^2_i\left(\sum_{j_1}^{2} (e^{(j_1)}_{i_2})^2 \left(\cdots \left( \sum_{j_{N-2}}^{2}(e^{(j_1j_2\cdots j_{N-2})}_{i_{N-1}})^2 \right)\cdots \right)\right) \right].\nonumber  \\
\end{eqnarray}
From Eqs.(\ref{e80}) and (\ref{e85}), we obtain

\begin{eqnarray}\label{e86}
 ||\mathcal{C}||^2-||\mathcal{C}\times_1 A_1\times_2 A_2^{(j_1)}||^2
& = & \frac{1}{2^N}\left[||\mathcal{T}||^2\right. \nonumber \\
& & \left. -\frac{1}{2^{N-2}}\sum_{i}^{3} c^2_i e^2_i \left(\sum_{j_1}^{2} (e^{(j_1)}_{i})^2 \left(\cdots \left( \sum_{j_{N-2}}^{2}(e^{(j_1j_2\cdots j_{N-2})}_{i})^2 \right)\right)\right) \right].\nonumber \\
\end{eqnarray}
We have $||\hat{e}||^2=e^2_1+e^2_2+e^2_3=1$ and $||\hat{e}^{(j_1\cdots j_{m-2})}||^2=(e_1^{(j_1\cdots j_{m-2})})^2+(e_2^{(j_1\cdots j_{m-2})})^2+(e_3^{(j_1\cdots j_{m-2})})^2=1,\; j_{m-2}=1,2,$ for $m=3,4,\cdots N.$ Let

\begin{equation}\label{e87}
c=\max\{|c_1|,|c_2|,|c_3|\}.
\end{equation}
Then, the second term in RHS of Eq.(\ref{e86}) is

\begin{eqnarray}\label{e88}
& &\frac{1}{2^{N-2}}\sum_{i}^{3} c^2_i e^2_i\left(\sum_{j_1}^{2} (e^{(j_1)}_{i_2})^2 \left(\cdots \left( \sum_{j_{N-2}}^{2}(e^{(j_1j_2\cdots j_{N-2})}_{i_{N-1}})^2 \right)\right)\right)\nonumber \\
& &\leq \frac{1}{2^{N-2}} |c|^2 \left[\sum_{i}^{3} e^2_i\left(\sum_{j_1}^{2} (e^{(j_1)}_{i_2})^2 \left(\cdots \left( \sum_{j_{N-2}}^{2}(e^{(j_1j_2\cdots j_{N-2})}_{i_{N-1}})^2 \right)\cdots \right)\right)\right]=c^2,
\end{eqnarray}
we can get the last equality in Eq.(\ref{e88}) by choosing appropriate vectors $\hat{e}$ and $\hat{e}^{(j_1j_2\cdots j_{k})}, \; j_k=1,2,$
for $k=1,2, \cdots N-2.$

Therefore, the minimization of Eq.(\ref{e86}) is by taking the maximization of the second term in RHS of it and the $GD_{A_1; A_2;\cdots; A_N}(\rho_{12\cdots N})$ is given by

\begin{equation*}
   GD_{A_1;A_2;\cdots; A_N}(\rho_{12\cdots N})=\frac{1}{2^N}\left[||\mathcal{T}||^2-c^2\right],
\end{equation*}
and $||\mathcal{T}||^2=c_1^2+c_2^2+c_3^2,$ hence the $GD$ is
\begin{equation}\label{e89}
 GD_{A_1;A_2;\cdots; A_N}(\rho_{12\cdots N})=\frac{1}{2^N}\left[ c_1^2+c_2^2+c_3^2-c^2\right],
 \end{equation}
which is as in ref. \cite{zhu}.

\section{Examples}
We apply our measure to the same multiqubit quantum states introduced in \cite{rad}.
The first example comprises the three-qubit mixed states,
\begin{equation}\label{e90}
  \rho = p |GHZ\rangle\langle GHZ|+\frac{(1-p)}{8} I_8,\; 0\leq p \leq 1;
\end{equation}
where $|GHZ \rangle =\frac{1}{\sqrt{2}}(|000\rangle+|111\rangle)$ and $I_8$ is the identity matrix.
Figure 1(a) shows the variation of $QD_{A_1;A_2;A_3}(\rho_{123})$ and $GD_{A_1;A_2;A_3}(\rho_{123})$ with $p$. We see that $GD_{A_1;A_2;A_3}(\rho_{123})$ increases continuously from the $p=0$ state (random mixture) to the $p=1$ state (pure GHZ state), as expected. For this state, it is known that entanglement is present only for $p > 1/5$ \cite{pit00, elt12}, showing quantum correlations can be present even when entanglement is zero and changing as similar to the $QD_{A_1; A_2; A_3}(\rho_{123})$ \cite{rad}. The $QD_{A_1;A_2;A_3}(\rho_{123})$ and $GD_{A_1;A_2;A_3}(\rho_{123})$ curves do not match, where the values of $GD_{A_1;A_2;A_3}(\rho_{123})$ {are} less than the values of $QD_{A_1;A_2;A_3}(\rho_{123})$ for $p$ between $0.1$ and $0.9$ approximately, as shown in Fig. 1(a).

The second example is the set of 3-qubit states.
\begin{equation}\label{e91}
\rho= p |W\rangle\langle W|+\frac{(1-p)}{8} I_8,\; 0\leq p \leq 1;
\end{equation}
where $|W\rangle=\frac{1}{\sqrt{3}}(|100\rangle+|010\rangle +|001\rangle)$. Figure 1(b) shows the variation of
$QD_{A_1;A_2;A_3}(\rho_{123})$ and $GD_{A_1;A_2;A_3}(\rho_{123})$ with $p$. The behaviours of $|W\rangle$ state in Eq.(\ref{e91}) is similar to the $|GHZ\rangle$ state in Eq.(\ref{e90}). The $QD_{A_1;A_2;A_3}(\rho_{123})$ and $GD_{A_1;A_2;A_3}(\rho_{123})$ show similar behavior, but the values of $GD_{A_1;A_2;A_3}(\rho_{123})$ is less than the values of $QD_{A_1;A_2;A_3}(\rho_{123})$ \cite{rad}, for all the values of $p > 0$ as shown in Fig.1(b).

In the last example, we consider the set of 3-qubit states
\begin{equation}\label{e92}
  \rho= p |000\rangle\langle 000|+(1-p) |+++\rangle\langle +++|,\; 0\leq p \leq 1;
\end{equation}
where $|+\rangle=\frac{1}{\sqrt{2}}(|0\rangle+|1\rangle)$. Figure 1(c) shows the variation of
$QD_{A_1;A_2;A_3}(\rho_{123})$ and $GD_{A_1;A_2;A_3}(\rho_{123})$ with $p$. The discord is symmetric about $p=\frac{1}{2}$ for both measures, at which they are maximum.
For $p=\frac{1}{2}$ the state can be written as $$\frac{1}{2}|000\rangle\langle 000|+\frac{1}{2}|+++\rangle\langle +++|$$
which is a mixture of classical states, which give non-zero discord. Our measure shows {similar behaviour} to the $QD_{A_1;A_2;A_3}(\rho_{123})$ \cite{rad}, but the values of $GD_{A_1;A_2;A_3}(\rho_{123})$ are less than the values of $QD_{A_1;A_2;A_3}(\rho_{123})$ for all the values of $p$ except for the small area around $p=\frac{1}{2}$, as shown in Fig. 1 (c).

\begin{figure}
  \centering
 \includegraphics[width=5 cm,height=4cm]{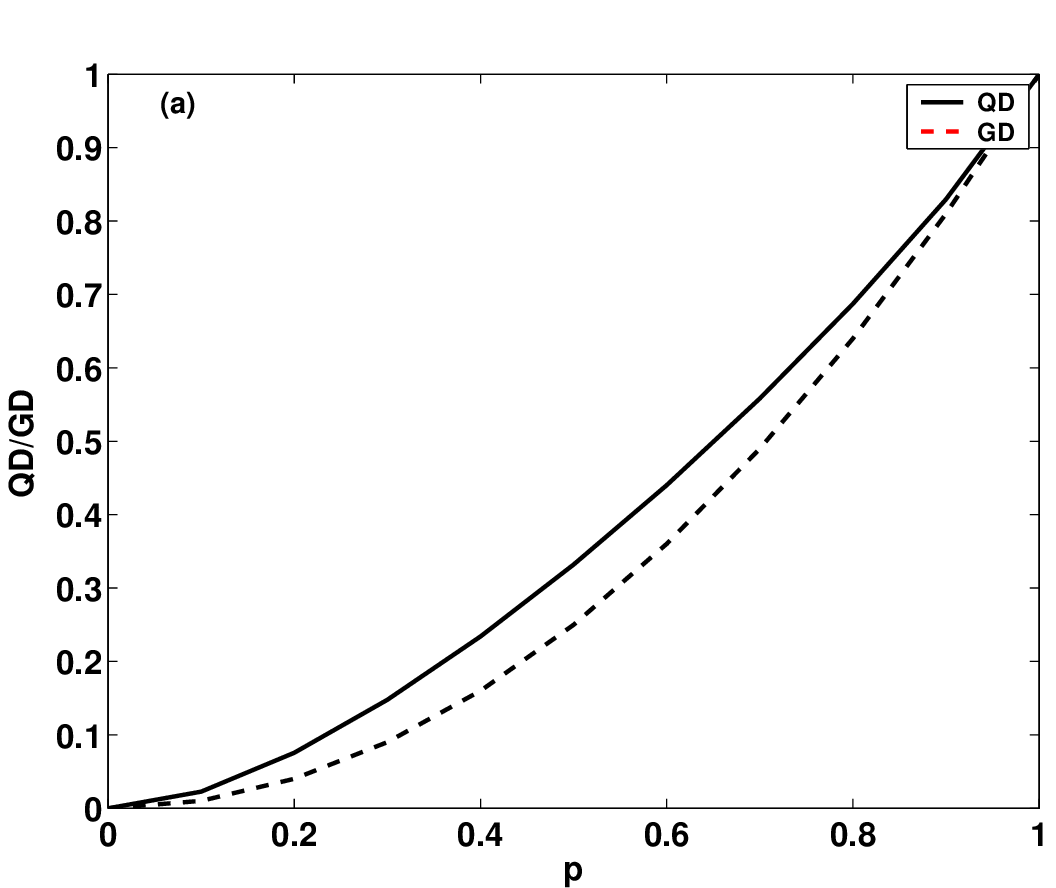}
 \includegraphics[width=5 cm,height=4cm]{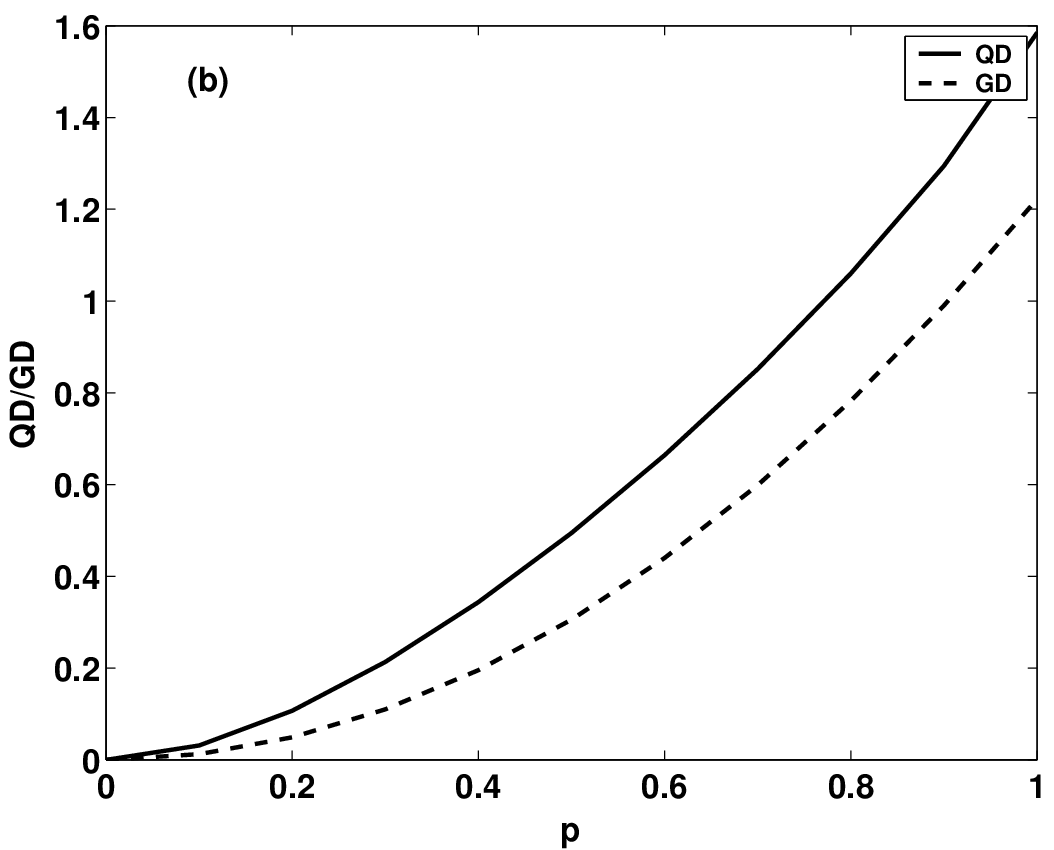}
 \includegraphics[width=5 cm,height=4cm]{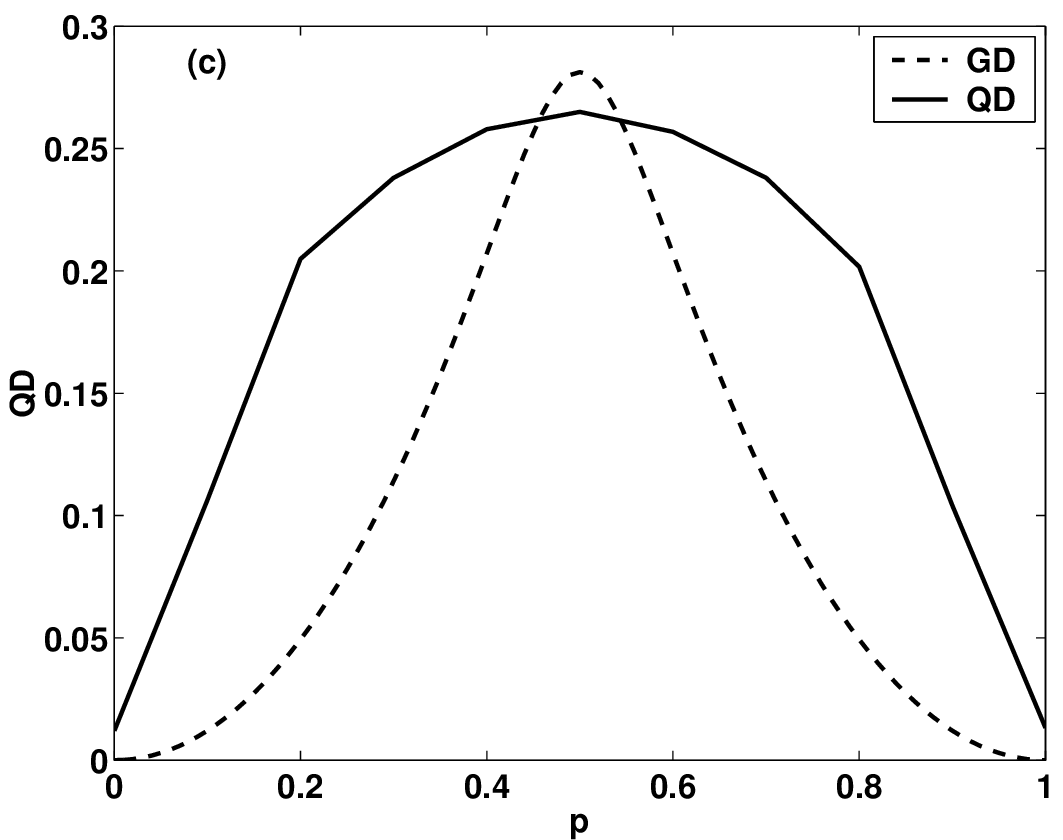}
 \caption{The tripartite measures of quantum discord. The states are: (a) Werner-GHZ states
$\rho = p |\psi\rangle\langle \psi|+\frac{(1-p)}{8} I_8,$
 where $|\psi\rangle = \frac{1}{\sqrt{2}}(|000\rangle+|111\rangle);$
(b) Werner-W states defined as $\rho$ with $|\psi\rangle = \frac{1}{\sqrt{3}}(|001\rangle+|010\rangle)+|100\rangle);$
(c) tripartite quantum correlated states $\rho= p |000\rangle \langle 000| +(1-p) |+++\rangle \langle +++|.$ }\label{1a}
\end{figure}

\section{Conclusion}

Radhakrishnan \emph{et. al} \cite{rad} proposed a generalization of quantum discord to multipartite systems, which is consistent with the conventional definition of discord in bipartite systems and derived explicit formulae
for any states. These results are significant in capturing quantum correlations for multi-qubit systems. In general,
the multipartite quantum discord is difficult to evaluate due to the complexity of its optimization.
We have presented a generic form of a generalization of the geometric measure of quantum discord for an $N$-partite state
using conditional measurements. We obtain the exact formulas of the measure for three, four, and $N$-qubit states.
The results in \cite{zhu} for the family of multi-qubit states {are reproduced} by our measure.

Future research will study the geometric quantum discord based on the trace distance to find computable analysis formulas for an arbitrary bipartite quantum state.  {Extending this} study to further geometric distance measures is another direction of research.\\

\noindent\textbf{{Acknowledgements}:}\\

We thank Fares A. Alzahri for his help with Latex and ﬁgures.

\end{document}